\documentclass[sigconf, nonacm]{acmart}
\usepackage{algorithm}
\usepackage{algpseudocode}
\usepackage{multirow}
\usepackage{subcaption}
\usepackage{enumitem}
\newcommand\vldbdoi{XX.XX/XXX.XX}
\newcommand\vldbpages{XXX-XXX}
\newcommand\vldbvolume{14}
\newcommand\vldbissue{1}
\newcommand\vldbyear{2020}
\newcommand\vldbauthors{\authors}
\newcommand\vldbtitle{\shorttitle} 
\newcommand\vldbavailabilityurl{https://github.com/kiarashgl/ExES}
\newcommand\vldbpagestyle{plain} 

\newcommand{\eat}[1]{}

\begin{document}

\title{Explaining Expert Search and Team Formation Systems with ExES}

\author{Kiarash Golzadeh}
\affiliation{%
  \institution{University of Waterloo}
}
\email{kiarash.golzadeh@uwaterloo.ca}

\author{Lukasz Golab}
\affiliation{%
  \institution{University of Waterloo}
}
\email{lgolab@uwaterloo.ca}

\author{Jaroslaw Szlichta}
\affiliation{%
  \institution{York University}
}
\email{szlichta@yorku.ca}

\begin{abstract}
Expert search and team formation systems operate on collaboration networks, with nodes representing individuals, labeled with their skills, and edges denoting collaboration relationships. Given a keyword query corresponding to the desired skills, these systems identify experts that best match the query. However, state-of-the-art solutions to this problem lack transparency. To address this issue, we propose ExES, a tool designed to explain expert search and team formation systems using factual and counterfactual methods from the field of explainable artificial intelligence (XAI).  ExES uses factual explanations to highlight important skills and collaborations, and counterfactual explanations to suggest new skills and collaborations to increase the likelihood of being identified as an expert.  Towards a practical deployment as an interactive explanation tool, we present and experimentally evaluate a suite of pruning strategies to speed up the explanation search.  In many cases, our pruning strategies make ExES an order of magnitude faster than exhaustive search, while still producing concise and actionable explanations. 
\end{abstract}

\maketitle

\pagestyle{\vldbpagestyle}

\eat{\begingroup\small\noindent\raggedright\textbf{PVLDB Reference Format:}\\
\vldbauthors. \vldbtitle. PVLDB, \vldbvolume(\vldbissue): \vldbpages, \vldbyear.\\
\href{https://doi.org/\vldbdoi}{doi:\vldbdoi}
\endgroup
\begingroup
\renewcommand\thefootnote{}\footnote{\noindent
This work is licensed under the Creative Commons BY-NC-ND 4.0 International License. Visit \url{https://creativecommons.org/licenses/by-nc-nd/4.0/} to view a copy of this license. For any use beyond those covered by this license, obtain permission by emailing \href{mailto:info@vldb.org}{info@vldb.org}. Copyright is held by the owner/author(s). Publication rights licensed to the VLDB Endowment. \\
\raggedright Proceedings of the VLDB Endowment, Vol. \vldbvolume, No. \vldbissue\ %
ISSN 2150-8097. \\
\href{https://doi.org/\vldbdoi}{doi:\vldbdoi} \\
}\addtocounter{footnote}{-1}\endgroup
}
\ifdefempty{\vldbavailabilityurl}{}{
\vspace{.3cm}
\begingroup\small\noindent\raggedright\textbf{Artifact Availability:}\\
The source code, data, and/or other artifacts have been made available at \url{\vldbavailabilityurl}.
\endgroup
}

\section{Introduction}

The \emph{expert search} problem \cite{husain2019expert} takes in a collaboration network, with nodes representing individuals, node labels corresponding to skills held by these individuals, and edges denoting collaboration relationships such as paper co-authorship. Given a keyword query consisting of the desired skills, the output is a ranked list of individuals who best match the query.  A related problem referred to as \emph{team formation} \cite{rad2022neural} is to identify a set of experts who collectively possess the desired skills and are closely connected in the collaboration network.  Expert search and team formation have been studied in various contexts, including 
academic  \cite{hamidi2023variational,xu2022academic}, 
medical \cite{medical}, 
social media influencers \cite{bozzon2013choosing}, 
human resources (including organizational expert search and talent recruitment) \cite{ha2015personalized,karimzadehgan2009enhancing,rostami2019t}, 
legal \cite{askari2022expert}, 
and question answering \cite{neshati2017dynamicity, ghasemi2021user}. 

Inspired by methods such as PageRank \cite{pagerank} and graph neural networks \cite{rad2022neural,hamidi2023variational}, state-of-the-art expert search and team formation systems combine a variety of signals from the collaboration network in non-obvious ways when ranking a given node: its skills, the skills of its collaborators and the network structure around it\footnote{Expertise can ``propagate": even if the node itself does not possess a skill, it may indirectly hold some expertise in, or be able to easily acquire it, if its collaborators have that skill \cite{serdyukov2008modeling, deng2012modeling}.}.
As a result, while these solutions are effective, they lack transparency, making them difficult to debug and limiting their practical uptake. 

The field of explainable artificial intelligence (XAI) was introduced to overcome the opacity of modern AI tools.  For example, \emph{factual} explanations assign importance scores to features, and \emph{counterfactual} methods perturb the features to change the model's output (e.g., a declined loan application would have been approved if the applicant's income were 20 percent higher) \cite{guidotti2019factual, bodria2023benchmarking}.  However, we are not aware of any post-hoc explanation methods for arbitrary expert search and team formation systems.

To fill this gap, we present ExES: a tool to explain expert search and team formation systems.  ExES is model-agnostic: it does not require access to the system's internal mechanisms and only needs to probe the system with different perturbations of the input and observe how the output changes.  Building on the ExES demo \cite{exesdemo}, we make the following contributions in this paper.  

\begin{enumerate}[nolistsep,leftmargin=*]
\item \textbf{Framework (Section~3.1).} 
We seek ways to leverage existing factual and counterfactual tools designed for classifiers.  To do so, we cast expert search and team formation as binary classification problems.  The features are the query and the collaboration network, and the class is whether a given individual is considered an expert or part of the team of experts. 
ExES can then use factual methods to identify features (e.g., skills or network connections) that were influential in expert selection, while counterfactual methods can support career advancement by finding skills and network connections that, when added, can turn non-experts into experts for a given query. 

\item \textbf{Scalability (Sections~3.2-3.5).} The space of candidate explanations spans all perturbations of the query and the collaboration network: query terms can be added or removed, node labels can be added or removed, edges can be added or removed, etc.  ExES exploits the structure of the expert search problem to prune this space by (1) focusing on network features in the neighborhood of the node being explained, (2) using beam search to limit the number of counterfactual explanations explored, (3), using word embeddings to guide the search for skill perturbations, and (4) using link prediction to guide the search for connection perturbations.  

\begin{figure*}[ht!]
    \centering
    \includegraphics[width=0.75\linewidth]{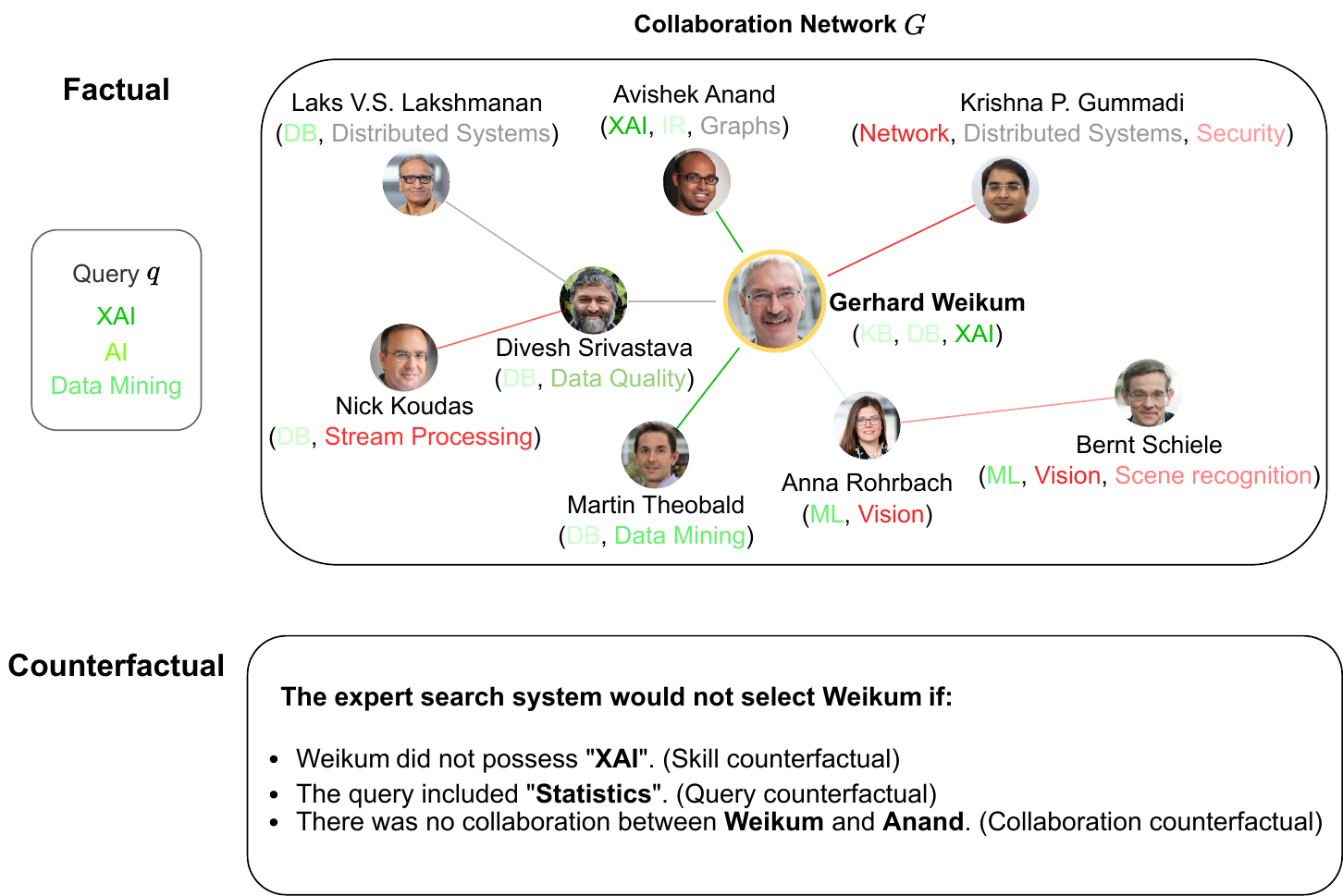}
    \caption{Factual and counterfactual explanations for $P_6$}
    \label{fig:toyexplanation}
\end{figure*}

\item \textbf{Evaluation (Section~\ref{sec:evaluation}).}  We assess the effectiveness and efficiency of ExES for explaining state-of-the-art expert search and team formation systems using two popular datasets: the DBLP academic collaboration network and the GitHub network.  We show that in many cases, our pruning strategies make ExES an order of magnitude faster than exhaustive search, while still producing concise and actionable explanations.

\end{enumerate}


As an example, Figure \ref{fig:toyexplanation} shows an academic collaboration network with eight researchers.   We issue a query for `XAI', `AI' and `Data Mining' to an expert search system $E$, which returns ``Gerhard Weikum" as the top expert.  Figure \ref{fig:toyexplanation} also shows the factual and counterfactual explanations generated by ExES.  For factual explanations, ExES colors the most important features in (dark) green and the least important features in (dark) red; recall that the feature space consists of the query and the entire collaboration network: the skills and connections of the individual being explained (Weikum) as well as the skills and connections of everyone else in the network.  We see that $E$ selected Weikum because his skills and his collaborators' skills match the query.  On the other hand, some of Weikum's collaborators have unrelated skills (in red), such as `Stream Processing,' and therefore these connections are also deemed unimportant (red edges).  Counterfactual explanations suggest that removing one of Weikum's skills or collaborations, or adding an unrelated keyword to the query such as `Statistics', would result in $E$ no longer ranking Weikum as the top expert.   

The remainder of this paper is organized as follows. Section~\ref{sec:relatedWork} reviews the related work, Section~\ref{sec:method} presents our solution, Section~\ref{sec:evaluation} contains our experimental evaluation, and Section~\ref{sec:conclusion} concludes the paper with directions for future work.

\section{Related Work} \label{sec:relatedWork}

\subsection{Expert Search}

As shown in Table \ref{tab:expertsearchrelatedwork}, we categorize expert search systems as document-based, graph-based and hybrid. Early solutions used statistical language models to calculate the probability that an expert, modelled as a document corpus (e.g., their publications), 
matches the given query \cite{balog1}.  These document-based solutions can be further divided into document-centric (retrieving documents relevant to the query, whose authors are then reported as experts \cite{momtazi1, macdonald1, Berger_2020}) and profile-centric (learning representation vectors for each individual and ranking them with respect to the query \cite{profile1, van2016unsupervised, balog2006formal,cifariello2019wiser}). Later approaches used word embeddings instead of exact word matching to recognize semantically similar terms  \cite{dargahinobari,fallahnejad2022attention}.  On the other hand, graph representation learning methods capture expertise propagation throughout the collaboration network \cite{deng2012modeling,brochier2020new, jimenez2022proficiencyrank, nikzad2021exem}. Popular approaches include calculating DeepWalk \cite{perozzi2014deepwalk}, PageRank \cite{pagerank}, and HITS \cite{kleinberg1999authoritative} scores to rank candidate experts for a given query.  Finally, recent hybrid methods combine textual and network signals to recommend experts \cite{nikzad2020berters, kang2023expfinder}.

\begin{table}
\small
    \centering
    \caption{A taxonomy of expert search solutions}
    \label{tab:expertsearchrelatedwork}
    \begin{tabular}{cc}
    \hline
        \textbf{Approach} & \textbf{References} \\
        \hline
        Document-based & \cite{balog1,momtazi1, macdonald1, Berger_2020, dargahinobari, profile1, van2016unsupervised, balog2006formal, fallahnejad2022attention, cifariello2019wiser} \\
        Graph Representation Learning & \cite{brochier2020new, jimenez2022proficiencyrank, nikzad2021exem} \\
         Hybrid Models & \cite{nikzad2020berters, kang2023expfinder} \\
         \hline
    \end{tabular}
    
\end{table}

\subsection{Team Formation}

In Table \ref{tab:teamformationsolutions}, we divide team formation solutions into graph optimization, integer programming, and neural.  Graph methods ensure that team members are close together in the collaboration network, e.g., by minimizing the distances among members \cite{kargar2011discovering,zihayatedbt,pagerank} or forming a minimum spanning tree \cite{lappas2009finding}.  Integer programming methods treat team formation as an optimization problem, e.g., facility location  \cite{neshati2014expert} or allocation of teams to tasks \cite{georgara2022allocating}.  Recent neural methods learn latent representations of individuals, e.g., using autoencoders \cite{sapienza2019deep,hamidi2021retrieving}, graph attention networks  \cite{kaw2023transfer}, or contrastive learning if a training dataset is available with examples of successful teams \cite{teamnegativesampling}. 

\begin{table}
\small
    \centering
    \caption{A taxonomy of team formation solutions}
    \label{tab:teamformationsolutions}
    \begin{tabular}{cc}
    \hline
        \textbf{Approach} & \textbf{References} \\
    \hline
         Graph Optimization & \cite{lappas2009finding, kargar2011discovering, zihayatedbt, pagerank} \\
         Integer Programming & \cite{neshati2014expert, georgara2022allocating} \\
         Neural Methods & \cite{sapienza2019deep, hamidi2021retrieving, teamnegativesampling, kaw2023transfer} \\
         \hline
    \end{tabular}   
\end{table}

We are aware of one prior work on explaining team formation systems \cite{georgara2022building}, but it is specific to integer programming solutions and only generates \textit{contrastive explanations}, which compare various characteristics (such as diversity, coherence, and member satisfaction) of selected teams with those of random teams. In contrast, ExES supports any expert search and team formation method, focusing on the impact of input features on the system's decision, even when team characteristics are unavailable. 

\subsection{Explainable AI}

There are several ways to characterize explainable AI methods. Intrinsic approaches call for models that are explainable by design, such as decision trees, whereas post-hoc approaches explain the outcomes of models that are already developed and trained.  Furthermore,  local methods explain individual predictions, while global methods provide an understanding of a model's overall behaviour and logic.  Finally, factual explanations, also known as feature-based or saliency methods, find the most important features for an individual prediction, and counterfactual explanations seek minimal perturbations to the input features to flip the model's prediction.  ExES is a post-hoc method to explain local outcomes of expert search and team formation methods, both factually and counterfactually, in a model-agnostic manner.  

Specifically, since expert and team search systems operate on collaboration graphs, we point out factual \cite{ying2019gnnexplainer,pgexplainer} and counterfactual \cite{lucic2022cf}  explanations for Graph Neural Networks (GNNs) that identify important graph elements or adjacency matrix perturbations in the context of predictions made by a GNN.  Moreover, since expert and team search systems usually rank their outputs according to relevance to the query, we also mention explanations of neural rankers, which (factually) identify important document fragments \cite{singh2019exs, chowdhury2023rank,lyu2023listwise} or (counterfactually) document and query perturbations that affect document ranking \cite{credence}. A novel aspect of ExES is that both graph structure and textual features are taken into account during the explanation, though our casting of expert search as a binary classification problem follows previous work on ranking explanations.

\section{Our Solution} 
\label{sec:method}


\subsection{Preliminaries and Solution Overview} 
\label{sec:preliminaries}

Table \ref{tab:notations} lists the symbols used in this section.
Let $S = \{s_1, s_2, \dots, s_l\}$ be the universe of skills, and $G = (P, E)$ be a node-labeled collaboration network, with individuals  ${P = \{p_1, p_2, \dots, p_n \}}$ as nodes, and connections ${E = \{(p_{i_1}, p_{j_1}), (p_{i_2}, p_{j_2}), \dots, (p_{i_m}, p_{j_m}) \}}$ between individuals as edges. Each $p_i$ possesses a set of skills $S_i$, for $S_i \subset S$.

Given a query $q$ consisting of a user-supplied set of skills ($q \subset S$)  and a value of $k$, the objective of  expert search is to identify the top $k$ experts that best match $q$. Let $\mathcal{R}_{p_i}(q, G)$ be the rank of expert $p_i$, with respect to query $q$ and the collaboration network $G$, produced by a solution $\mathcal{R}$ to the expert search problem.

Furthermore, the goal of team formation is to find a subset of nodes from $G$, such as ${T = \{p_{T_1}, p_{T_2}, \dots, p_{T_k}\}} \subset P$, so that $q \subset \bigcup_{i=1}^{k}S_{T_i}$. We refer to the team formation system as $\mathcal{F}$, where $\mathcal{F}(q, G)$ is the resulting team given the query $q$ and the collaboration network $G$.

\begin{table}[t]
\small
    \centering
        \caption{Important symbols}
    \label{tab:notations}
    \begin{tabular}{cc}
    \hline
         \textbf{Notation} & \textbf{Description} \\
         \hline
         $S$ & Universal set of skills \\
         $S_i$ & Skill set of individual $p_i$ \\
         $G$ & Collaboration Network \\
         $\mathcal{R}_{p_i}(q, G)$& Rank of $p_i$ wrt query $q$ and network $G$ \\
         $\mathcal{C}_{p_i}(q, G)$& Relevance status of $p_i$ wrt query $q$ and network $G$ \\
         $\mathcal{F}(q, G)$& Team formed for query $q$ and network $G$  \\
         $\mathcal{M}_{p_i}(q, G)$& Membership status of $p_i$ wrt query $q$ and network $G$  \\
         $\mathcal{N}(p_i)$& Neighborhood of $p_i$ \\
         $S_{\mathcal{N}(p_i)}$& Skills included in the neighborhood of $p_i$\\
         $d$ & Neighborhood radius \\
         $b$ & Beam size \\
         $t$ & Number of candidate features \\
         $\gamma$ & Maximum explanation size\\
         \hline
    \end{tabular}

\end{table}


The goal of ExES is to explain the decision-making process behind $\mathcal{R}$ or $\mathcal{F}$.  To leverage existing factual and counterfactual tools for feature-oriented post-hoc explanations, we need to formulate our problems in terms of input features and an output decision.  The features are the skills requested in the query, the skills held by each node, and the edges in the collaboration network.  As for the output decision, we cast our problems as  
binary classification problems, inspired by prior work in explainable information retrieval \cite{singh2019exs,credence}.  Given a query $q$ against a collaboration graph $G$, we ask, for a given node $p_i$, whether $\mathcal{R}_{p_i}(q, G) \leq k$.  Let  $\mathcal{C}_{p_i}(q, G)$ be the resulting \emph{relevance status}, true if $p_i$ was deemed to be an expert (i.e., ranked inside the top-$k$) and false otherwise.  Similarly, in the context of team formation, let $\mathcal{M}_{p_i}(q, G)$ be the \emph{membership status} of $p_i$, true if  $p_i \in \mathcal{F}(q, G)$ and false otherwise.


\begin{figure}[t]
    \centering
    \includegraphics[width=0.99\linewidth]{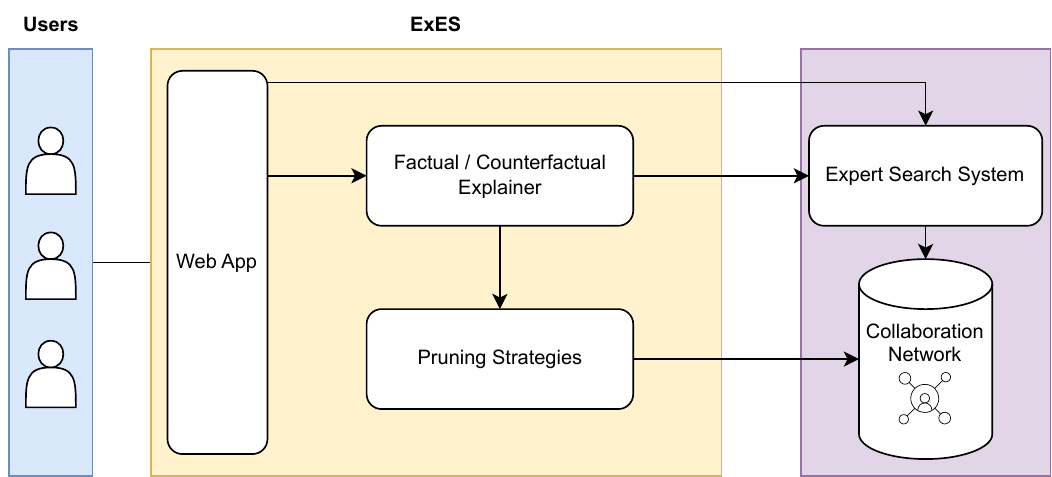}
    \caption{ExES Architecture}
    \label{fig:arch}
    \vskip -0.18in
\end{figure} 

Figure \ref{fig:arch} shows the ExES workflow and architecture.  Given an expert search or team formation system that is to be explained, the user issues a keyword query, representing the desired skills, and the system outputs a ranked list or a team of experts.  The user selects an individual from the collaboration graph whose presence or absence in the output is to be explained.  ExES then probes the underlying system with perturbed inputs to generate factual and counterfactual explanations.  Along the way, ExES uses a suite of pruning strategies that exploit our problem structure to speed up the explanation search. 

Starting with expert search, we discuss factual explanations in Section~\ref{saliency}, counterfactual explanations in Section~\ref{counterfactual}, computational complexity with and without pruning in Section~\ref{sec:complexity}, and extensions to team formation in Section~\ref{teamformation}.  We include examples of various types of explanations, illustrated with screenshots from ExES.

\subsection{Factual Explanations} \label{saliency}

For factual explanations, ExES is compatible with any feature scoring method.  In our prototype, we use SHAP
\cite{lundberg2017shap}, a popular method that computes a contribution score for each feature, by making multiple inference calls to the underlying model with different values of that feature.  
The higher the score, the more likely it is that a change to this feature changes the model's prediction, which in our case is the relevance status $\mathcal{C}_{p_i}(q, G)$ or the membership status $\mathcal{M}_{p_i}(q, G)$. 

To explain the relevance status of node $p_i$ (with respect to some query) using SHAP, a trivial approach is to find the SHAP values for all input features, i.e., every query keyword, every skill assigned to every node, and every edge in the collaboration network.  When confronted with large real-world collaboration networks, we require pruning methods to deliver explanations to users at interactive speeds.  In the remainder of this section, we introduce several pruning methods informed by the structure of our problem.

\textbf{Pruning Strategy 1: Network Locality.} For a given query, we posit that $p_i$'s relevance status is most strongly influenced by the query keywords, by its own skills, and by and the network structure (collaborators and their skills) in its \emph{neighborhood}, defined as the induced subgraph of nodes located within a distance threshold $d$ from $p_i$ (see Section  \ref{sec:parametersensitivityanalysis} for a parameter sensitivity analysis). We refer to the neighborhood of $p_i$ as $\mathcal{N}(p_i)$, and denote the skills included in the neighborhood as $S_{\mathcal{N}(p_i)}$.  Thus, when it comes to skills, the features that will be scored by SHAP are only the skills mentioned in the query and the skills in $S_{\mathcal{N}(p_i)}$.   

\begin{figure}
    \centering
    \includegraphics[width=\linewidth]{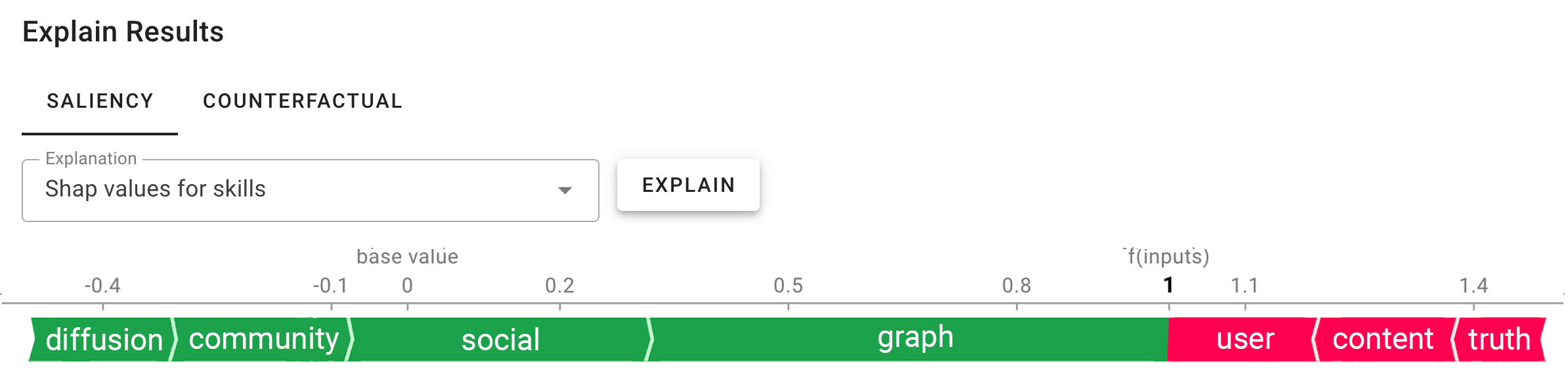}
    \caption{Skill factual explanations}
    \label{fig:shapskill}
\end{figure}

\textbf{Example 1.} Consider the collaboration network obtained from DBLP (dataset details in Section~\ref{subsec:setup}). Given the query ``social graph'' and $k=10$, suppose we want to explain the relevance status of expert ``Jure Leskovec'', ranked within the top ten. Figure \ref{fig:shapskill} displays the SHAP values of Leskovec's skills in a \textit{force plot}, with important skills in green and interval lengths corresponding to SHAP scores.  The \textit{base value} of zero on ther x-axis shows the expected value of the relevance status and $f(inputs)$ corresponds to Leskovec's relevance status of one.  Leskovec possesses both of the query keywords, ``social'' and ''graph'', both of which have the highest SHAP scores.  His skills unrelated to the query, such as ``user'', are shown in red.  

Likewise, we only consider edges within a distance threshold $d$ of $p_i$.  To further limit the number of edges whose SHAP values will be computed, we focus on \emph{influential} collaborations.

\textbf{Pruning Strategy 2: Influential Collaborations.}
We initialize a queue $\mathcal{Q}$ of \textit{impactful experts} with $p_i$, and an empty set $\mathcal{I}$ of \textit{impactful links}. Starting from $p_i$, in each iteration, we expand the first unexpanded impactful expert from $\mathcal{Q}$, denoted as $p_x$, and calculate the SHAP values of its incident edges. For edges $(p_x, p_y)$ with absolute SHAP values beyond a specified threshold $\tau$, we add them to $\mathcal{I}$, and append $p_y$ to the end of $\mathcal{Q}$. This threshold limits the branching factor at each stage. In the end, we calculate the SHAP values of only the potentially impactful links in $\mathcal{I}$.

\begin{figure}
    \includegraphics[width=\linewidth]{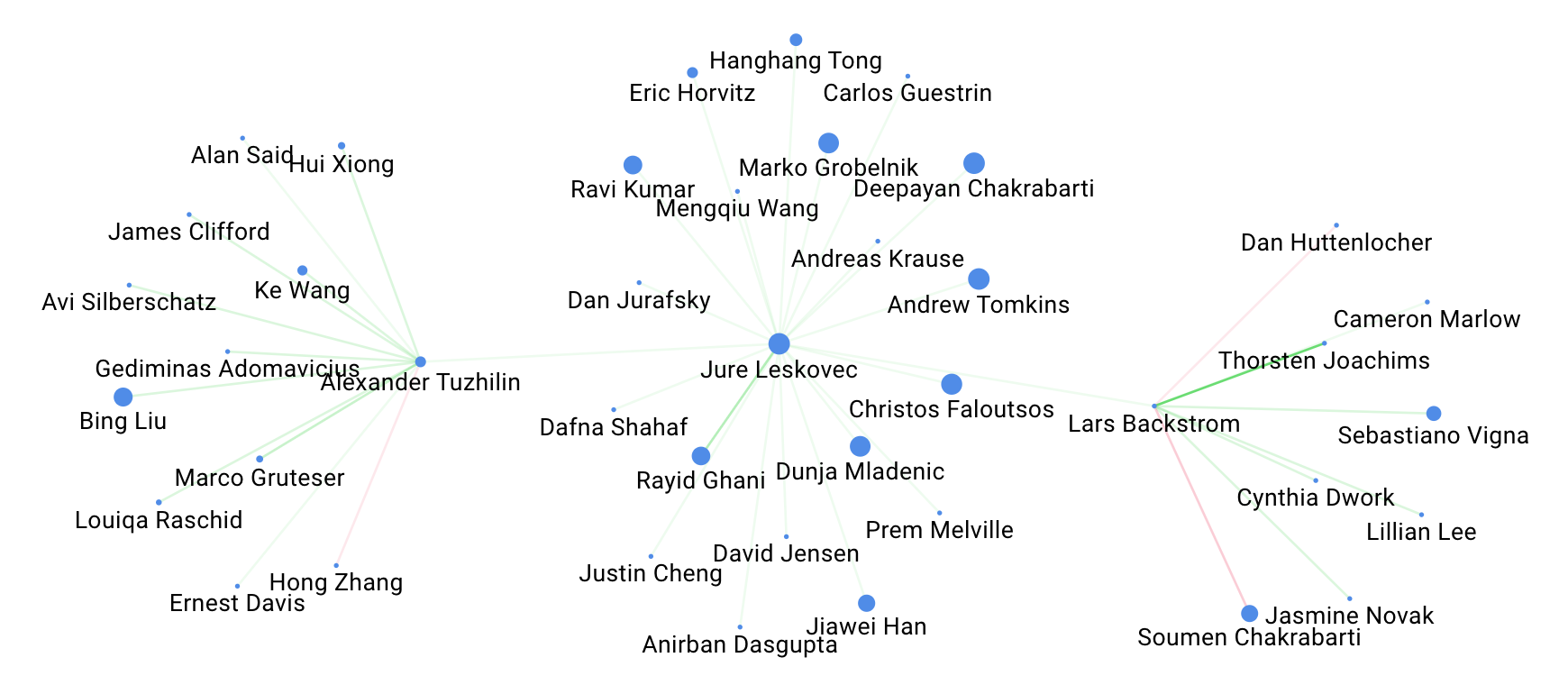}
    \caption{Collaboration factual explanations}
    \label{fig:shapgraph}
\end{figure}

\textbf{Example 2.}
Returning to Jure Leskovec from Example 1, Figure \ref{fig:shapgraph} shows the SHAP values of Leskovec's connections (only those selected by the pruning rules) using a node-link diagram, where green edges show a positive effect and red edges show a negative effect toward the relevance status. Edge opacity indicates the degree of importance of each edge, and the size of the nodes shows the rank of the corresponding node with respect to this query. The higher the expert is in the ranking, the larger the node is in the plot. This plot shows that while some of Leskovec's neighbors, like Prem Melville (located at the bottom of the middle part of the figure), are not highly ranked with respect to the given query, their presence still contributes to Leskovec's relevance for the query.

\subsection{Counterfactual Explanations}\label{counterfactual}

To explain the relevance status of a node $p_i$ counterfactually, we perturb the features. i.e., the search query $q$ and the collaboration network $G$, to flip $\mathcal{C}_{p_i}(q, G)$.  That is, we identify changes that would turn experts (ranked in the top-$k$) into non-experts (ranked outside the top-$k$) and vice versa.  
Following previous work in XAI, we seek \emph{minimal} explanations in terms of the number of features modified, which are likely to be more actionable (e.g., the fewest new skills one would have to acquire to be a highly-ranked expert for a given query).  As in factual explanations, we introduce search space pruning rules to speed up explanation generation.

\textbf{Pruning Strategy 3: Beam Search.} Shown in Algorithm \ref{alg:cap}, we use this strategy for all types of counterfactual explanations.  We use beam width $b$ to consider perturbations up to a maximum size (i.e., number of features perturbed) $\gamma$. The search continues until it finds $e$ minimal explanations within this space.

In line \ref{line:1},  we select $t$ candidate features to include in the beam search, guided by additional pruning strategies that depend on the type of explanation (details to follow). 
%
%
Inside the while loop (line \ref{line:5}), the beam search algorithm expands the perturbation states in $queue$ until $e$ explanations are found or the queue is empty. Each iteration begins by initializing $expandedQueue$, which contains the expanded states, with an empty set. Then, we expand every state in $queue$ by appending each candidate perturbation to it (line \ref{line:9}). After applying the perturbations to $G$ or $q$ (line \ref{line:10}), we compute the new rank and relevance status (lines \ref{line:11}, \ref{line:12}).  Counterfactual perturbations (i.e., those that change the relevance status) are added to $\mathcal{E}$ (lines \ref{line:13}, \ref{line:14}). Furthermore, perturbations not exceeding $\gamma$ are queued in $expandedQueue$, along with their new rank, for further expansion (lines \ref{line:16} and \ref{line:17}).

After this, the top $b$ states are selected from $expandedQueue$ based on the new rank (line \ref{line:21}), and passed to $queue$ for the next iteration (lines \ref{line:23} and \ref{line:24}). The sorting direction is determined based on the initial relevance; i.e. if the initial relevance is 1, the sorting direction is descending, otherwise it is ascending. After the search has ended, the explanations in $\mathcal{E}$ are returned (line \ref{line:27}).

\renewcommand{\algorithmicrequire}{\textbf{Input:}}
\renewcommand{\algorithmicensure}{\textbf{Output:}}
\begin{algorithm}[t]
\caption{Generating counterfactual explanations}\label{alg:cap}
\begin{algorithmic}[1]
\Require Collaboration network $G$, Query $q$, Ranker $\mathcal{R}$, Relevance status function $\mathcal{C}$, Word embedding $W$, Link prediction model $L$, expert $p_i$, Number of explanations $e \ge 1$, Number of candidate features $t \ge 1$, Beam width $b$, Maximum perturbation size $\gamma$
\Ensure List of $e$ explanations $\mathcal{E}$
\State $candidateFeatures \gets \Call{getCandidateFeatures}{t, G, W, L}$ \label{line:1}
\State $\mathcal{E} \gets \emptyset$ \label{line:2}
\State $queue \gets \{\emptyset\}$ \label{line:3}
\State $initialRelevance \gets \mathcal{C}_{p_i}(q, G)$\label{line:4}
\While{$|\mathcal{E}| < e$ and $|queue| > 0$} \label{line:5}
    \State $expandedQueue \gets \emptyset$\label{line:6}
    \ForAll{$perturbation \in queue$}\label{line:7}
        \ForAll{$feature \in candidateFeatures$}\label{line:8}
            \State $expandedPerturbation \gets perturbation \cup \{feature\}$\label{line:9}
            \State $G', q' \gets \Call{Apply}{perturbation, G, q}$\label{line:10}
            \State $newRank \gets \mathcal{R}_{p_i}(q', G')$\label{line:11}
            \State $newRelevance \gets \mathcal{C}_{p_i}(q', G')$\label{line:12}
            \If{$newRelevance \ne initialRelevance$}\label{line:13}
                \State $\mathcal{E} \gets \mathcal{E} \cup expandedPerturbation$\label{line:14}
            \EndIf
            \If{ $|expandedPerturbation| < \gamma$} \label{line:16}
                \State $expandedQueue \gets expandedQueue \cup \{\langle newRank, expandedPerturbation \rangle\}$ \label{line:17}
            \EndIf
        \EndFor
    \EndFor
    \State $expandedQueue \gets \Call{selectTopK}{expandedQueue, b}$ \label{line:21}
    \State $queue \gets \emptyset$ \label{line:22}
    \For{$i \in [0, b)$} \label{line:23}
        \State $queue \gets queue \cup \{expandedQueue[i][1]\}$ \label{line:24}
    \EndFor
\EndWhile
\State \Return $\mathcal{E}$\label{line:27}
\end{algorithmic}
\end{algorithm}

We now present the details of line 1 (\textsc{GetCandidateFeatures}) for counterfactual skill, query and collaboration explanations, leading to further pruning strategies.

\subsubsection{Counterfactual Skill Explanations}\label{counterfactualskill}

Here, we consider adding or removing skills.  As in factual explanations, to reduce the space of possible counterfactuals, we do so only for $p_i$ or the collaborators in $p_i$'s neighborhood (recall Pruning Strategy 1). Given a query $q$, adding skills that are in $q$ (or similar) to $p_i$ or its neighborhood collaborators should move $p_i$ up in the ranking, and removing such skills should do the opposite.

\textbf{Pruning Strategy 4: Word Embeddings.} After pruning the number of nodes whose skills will be counterfactually perturbed, the next question is: which skills to add or remove?  To limit the space of skill counterfactuals, we train a word embedding model, denoted $W$, such as Word2Vec \cite{word2vec},  
on the textual expertise corpus from which the collaboration network labels were extracted. 
%
%
%
%
%
When adding skills to turn non-experts into experts, we select the $t$ most similar skills to the query 
based on $W$.  To turn experts into non-experts, we remove skills in $S_{\mathcal{N}(p_i)}$ that are the most similar to the query.

\begin{figure}[t]
    \centering
        \includegraphics[width=\linewidth]{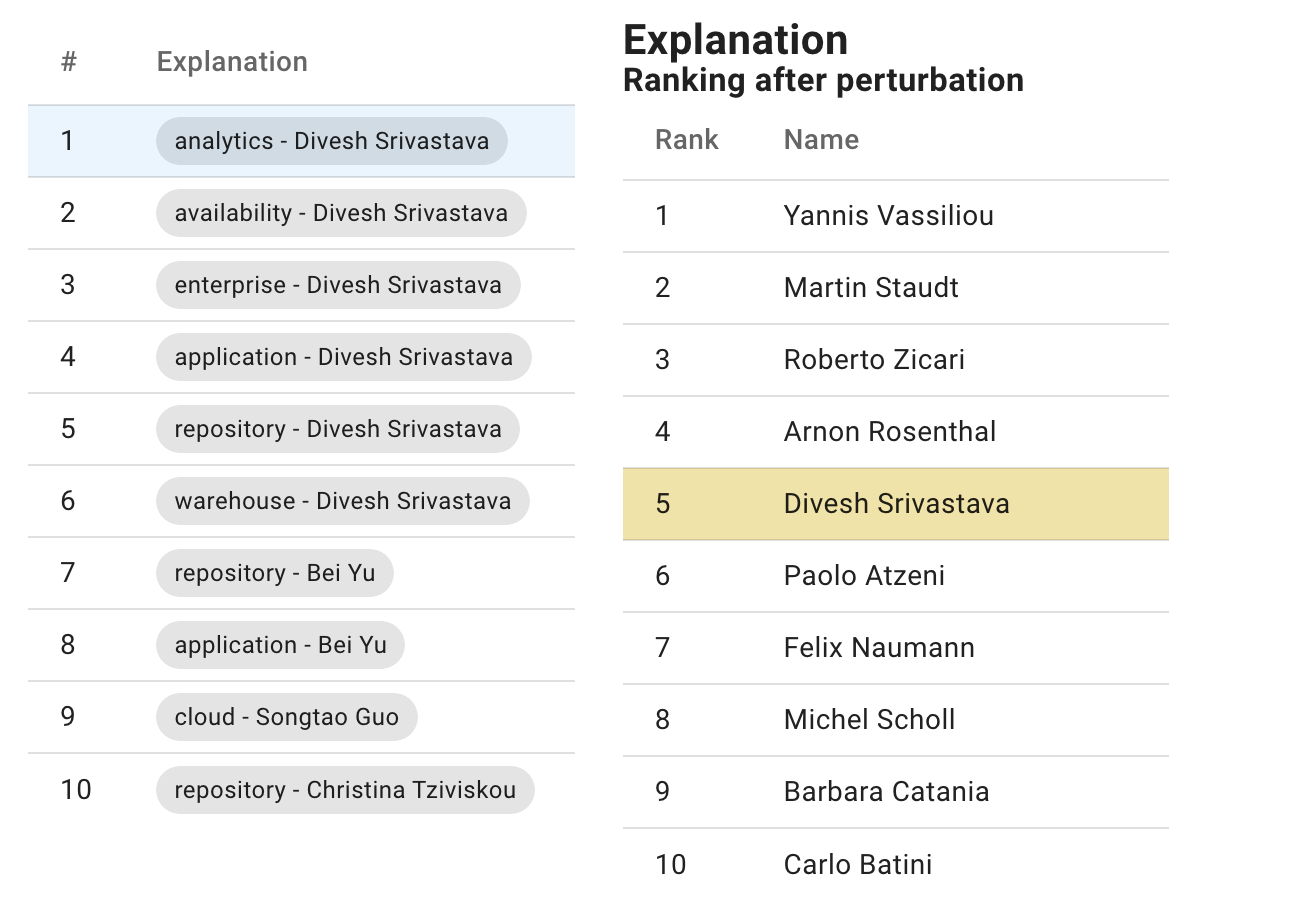}
        \caption{Counterfactual skill explanations}
        \label{fig:addskill}
\end{figure}

\textbf{Example 3.} Consider the query ``database management quality'' on the DBLP collaboration network with $k=10$, where  ``Divesh Srivastava'' is ranked 11th, i.e., outside the top-$k$. Suppose we want to counterfactually explain why Srivastava was ranked outside the top-10. Figure~\ref{fig:addskill} shows a list of skill addition explanations. These skills are either added to the target individual (Srivastava), or their neighbors (Bei Yu, Songtao Guo, Christina Tziviskou).  As an example, adding the \textit{analytics} skill improves Srivastava's rank for this query to fifth. The explanations are sorted by their size, i.e., the number of skills to add.  Explanations with the same size are further sorted by their effect on the ranking.  

\subsubsection{Counterfactual Query Explanations} \label{sec:counterfactualquery}

To produce counterfactual explanations for the relevance status of a node 
$p_i$ in terms of the search query, we apply query augmentation \cite{queryaugmentation,queryaugmentation2}.  We note that expert search queries usually consists of a small number of keywords, so perturbations via keyword removal may not be meaningful.  Therefore, we only consider keyword addition, where
adding skills from $S_i$ into $q$ should improve $p_i$'s ranking, and adding skills unrelated to $S_i$ into $q$ should do the opposite.


To find query augmentations that improve the ranking of $p_i$ to bring $p_i$ into the top-$k$, we again prune the skill space by identifying $t$ most similar skill keywords to the expert skill set $S_i$ and the query $q$, according to the keyword embedding model $W$ (recall Pruning Strategy 4). 
To find counterfactual query explanations that evict an expert $p_i$ from the top-$k$ list, we select skills similar to the query but different from $S_i$.

\subsubsection{Counterfactual Collaboration Explanations} \label{sec:counterfactualcollaboration}

Finally, to create counterfactual explanations 
in terms of collaborations, we add or delete connections in $p_i$'s neighborhood.  Intuitively, $p_i$'s ranking can improve if we add an edge 
to an expert for the given query $q$, and vice versa. 

 
 


\textbf{Pruning Strategy 5: Link Prediction.} To decide which connections to add in $p_i$'s neighborhood, we use a link prediction model, denoted $L$, trained on the collaboration network connections.  In our implementation, we use the Graph Auto-encoder (GAE) \cite{kipf2016variational} due to its high link prediction accuracy \cite{li2024evaluating}. 
$L$ serves as a recommender for new collaborations, to eliminate less promising collaborations from the search space of counterfactuals.  
To improve $p_i$'s ranking, we locate the most likely candidates for future collaborations within $\mathcal{N}(p_i)$, using the link prediction model $L$. 
To identify minimal sets for edge removal, we focus on edges in $\mathcal{N}(p_i)$ whose elimination worsens $p_i$'s ranking the most. 

\begin{figure}
\centering
    \includegraphics[width=\linewidth]{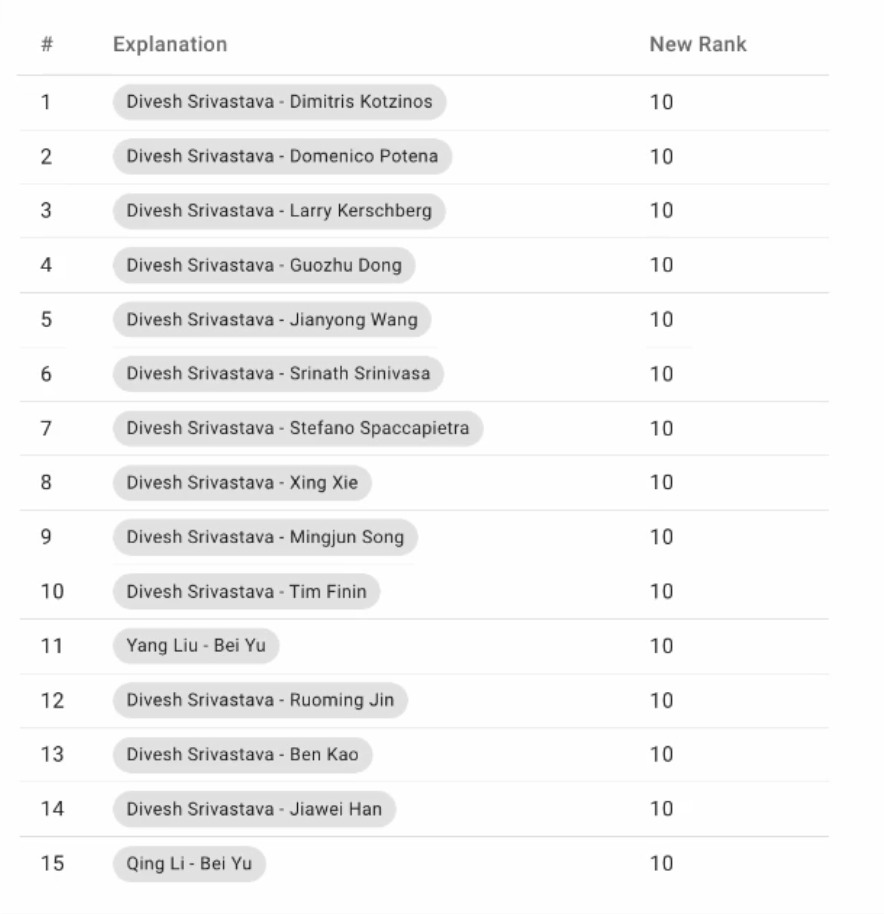}
    \caption{Counterfactual collaboration explanations}
    \label{fig:alledges}
\end{figure}


\textbf{Example 4.} Recall example 3, with the query ``database management quality'' on the DBLP collaboration network, with $k=10$, and ``Divesh Srivastava'' ranked outside the top-10. Figure \ref{fig:alledges} shows a list of counterfactual collaboration additions that place Srivastava inside the top-10. 

\subsection{Time Complexity} \label{sec:complexity}

\subsubsection{Factual Explanations}

According to \cite{lundberg2017shap}, the time complexity of calculating SHAP values of a prediction with $M$ input features is $O(2^M \times T)$, where $T$ is the inference time of the underlying model. Table \ref{tab:saliencytime} shows the complexity of the factual explanations supported by ExES, with and without pruning, where $T_{ranking}$ is the time complexity of the underlying expert search system for one query. 

When computing skill SHAP values to explain the relevance status of $p_i$, the input includes all skills held by individuals in $G$, a total of $\sum_{i=1}^n|S_i|$ features. In the worst case, this equals $|P| \times |S|$. However, ExES limits the features for SHAP value computation by only considering the skills of individuals within $p_i$'s neighborhood (Pruning Strategy 1). This reduces the number of features for SHAP to $\sum_{x|p_x\in \mathcal{N}(p_i)}|S_x| = |S_{\mathcal{N}(p_i)}|$. Similarly, to calculate SHAP values of collaborations in $G$, the features comprise every edge in $G$, for a total of $|E|$. ExES limits the input features by only including edges within the neighborhood of $p_i$, $\mathcal{N}(p_i)$. The number of edges selected during bread-first search is at most $|\mathcal{N}(p_i)|$, which is significantly smaller than $|E|$. ExES does not apply any pruning method for calculating SHAP values of query keywords. Thus, the time complexity is the same for ExES and exhaustive search.

\begin{table}
\footnotesize
    \centering
    \caption{Time complexity of factual explanations}
    \label{tab:saliencytime}
    \begin{tabular}{ccc}
    \hline
    \multirow{2}{*}{\textbf{Type}} & \multicolumn{2}{c}{\textbf{Time Complexity}} \\
    & \textbf{with Pruning} & \textbf{without Pruning} \\
    \hline
        skills & $O(2^{|S_{\mathcal{N}(p_i)}|}\times T_{ranking})$ & $O(2^{|P|\times|S|}\times T_{ranking})$\\
          query & $O(2^{|q|}\times T_{ranking})$ & $O(2^{|q|}\times T_{ranking})$ \\
         collaborations& $O(2^{|\mathcal{N}(p_i)|}\times T_{ranking})$ & $O(2^{|E|}\times T_{ranking})$ \ \\
    \hline
    \end{tabular}
    
\end{table}
\subsubsection{Counterfactual Explanations}
\begin{table}
\footnotesize
    \centering
    \caption{Time complexity of counterfactual explanations}
    \label{tab:counterfactualtime}
    \begin{tabular}{ccc}
    \hline
    \multirow{2}{*}{\textbf{Type}} & \multicolumn{2}{c}{\textbf{Time Complexity}} \\
    & \textbf{with Pruning} & \textbf{without Pruning} \\
    \hline
        skills & $O(T_W + b t  \gamma\times |\mathcal{N}(p_i)| \times T_{ranking})$&  $ O(2^{|S| \times |P|} \times  T_{ranking})$ \\
          query & $O(T_W + b t \gamma\times T_{ranking})$& $O(2^{|S|}\times T_{ranking})$  \\
         collaborations & $O(T_L + b t \gamma\times |\mathcal{N}(p_i)| \times T_{ranking})$& $O(2^{|E|}  \times T_{ranking})$ \\
    \hline
    \end{tabular}

\end{table}

Table \ref{tab:counterfactualtime} summarizes the complexity of counterfactual explanations, with and without pruning.  $T_W$ is the running time of the keyword embedding model to identify $t$ potential keywords, and $T_L$ is the runtime of the link prediction model to find $t$ potential edges.  

Exhaustive search calculates the new ranking and relevance status for every subset of the search space (ordered by the perturbation size) and then selects the minimal perturbations that flip the relevance status. For skill addition counterfactuals, the size of the search space is the count of missing skill-individual pairs in $G$, which is 
$\sum_{i \in P} |S - S_i|$. Conversely, for skill removal counterfactuals, it is the number of existing skill-individual pairs, which is $\sum_{i \in P} |S_i|$. These numbers could be equal to $|S| \times |P|$ in the worst case.
For query counterfactuals, the search space contains every missing keyword from the query, with a maximum size of $|S - q|$.
For collaboration addition counterfactuals, the search space contains every missing edge, of which there are $\binom{n}{2} - |E|$, and for collaboration removal counterfactuals, it includes all  $|E|$ edges. 

On the other hand, as stated in Section \ref{counterfactual}, ExES uses beam search for generating counterfactual explanations (Pruning Strategy 3). Recall that $b$ is the beam size and $\gamma$ is the maximum explanation size. According to \cite{russell2010artificial}, the time complexity of beam search with beam size $b$, branching factor $w$, and maximum depth $\gamma$ is $O(b \times w \times \gamma)$. For skill counterfactuals, ExES limits the search to the expert's neighborhood $\mathcal{N}(p_i)$ and selects $t$ keywords as potential skill perturbations, making the search space and branching factor $t\times \mathcal{N}(p_i)$. For query counterfactuals, the search space and branching factor is $t$, corresponding to the $t$ keywords added to the query.
For collaboration counterfactuals, ExES narrows down the potential added/or removed edges to manage the search space, selecting $t$ perturbed edges for both additions and removals involving $\mathcal{N}(p_i)$, keeping the search space and branching factor at $t$. Thus, pruning the search space reduces the running time from exponential to polynomial. 

\subsection{Explaining Team Formation Systems} 
\label{teamformation}

The explanations and pruning strategies described in Sections \ref{saliency} and \ref{counterfactual} apply to team formation as well -- it suffices to replace relevance status with team membership as the binary class label.  ExES can output SHAP scores for why an expert was included (or not) in a team, and counterfactual explanations for what would have to change to exclude (or include) an expert from a team.  The time complexity of team formation explanations is similar to that shown in Section \ref{sec:complexity}, with 
$T_{ranking}$ substituted with $T_{teamFormation}$, the time complexity of executing the team formation procedure for a single pass.

\begin{figure}[t]
    \centering
    \includegraphics[width=0.99\linewidth]{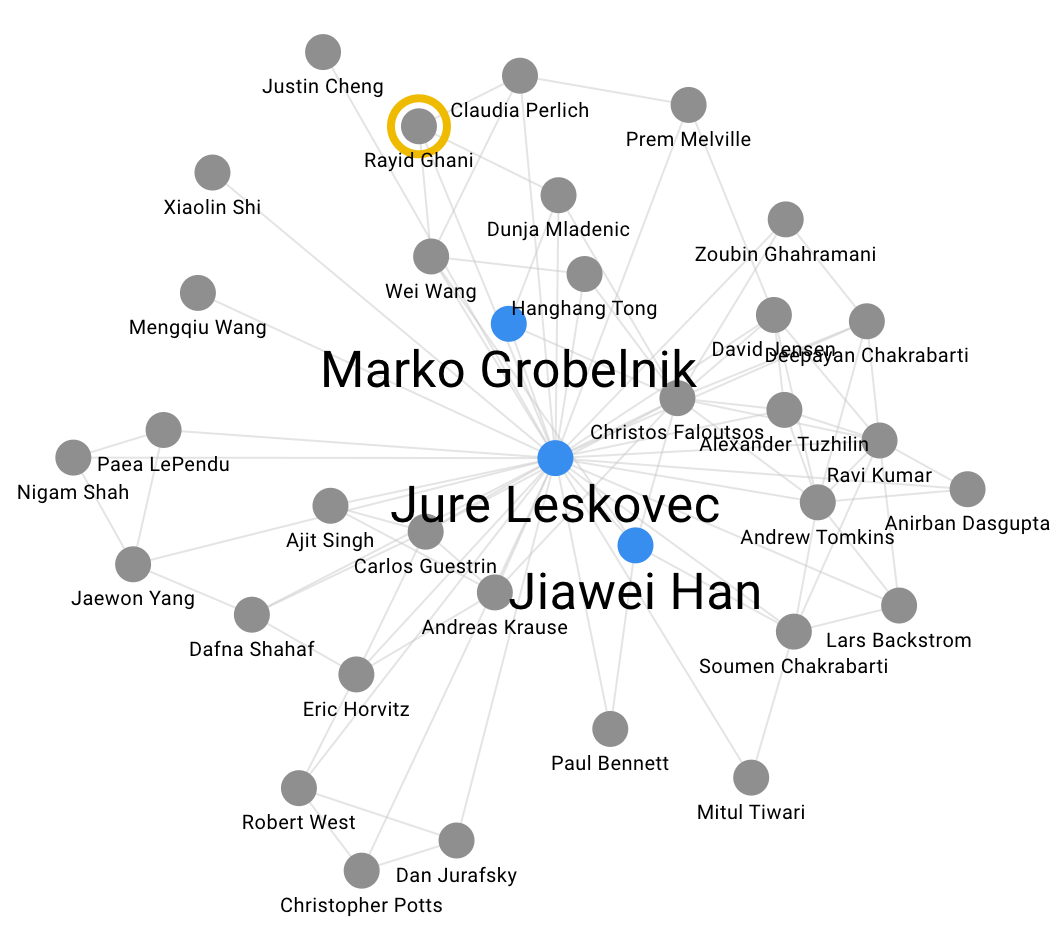}
    \caption{A team of experts}
    \label{fig:teamformation1}
\end{figure}

\begin{figure}[t]
    \centering
    \includegraphics[width=\linewidth]{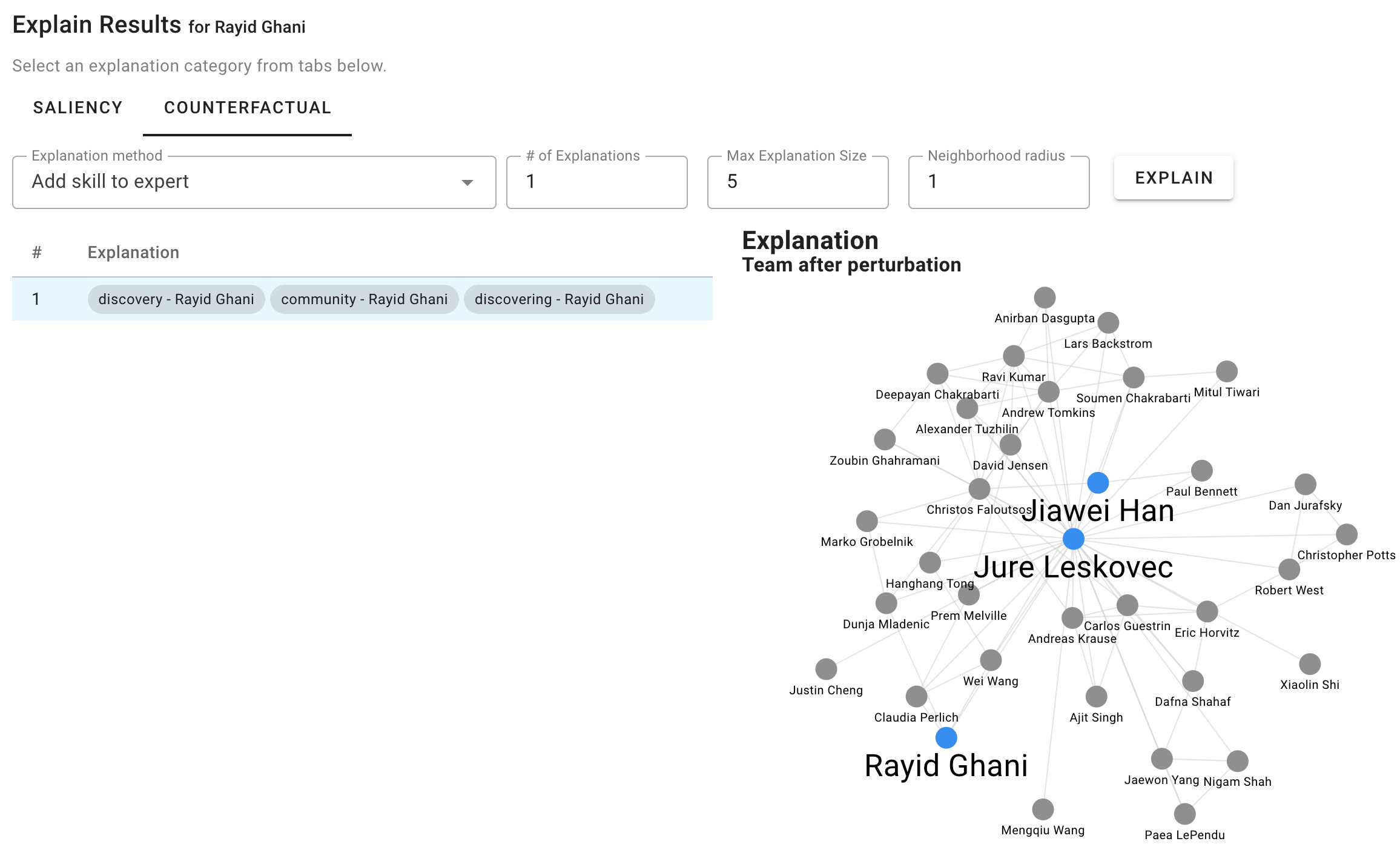}
    \caption{A counterfactual explanation for excluding an expert from the team}
    \label{fig:teamformation2}
\end{figure}

\textbf{Example 5.} Consider the query ``social graph pattern mining" on the DBLP collaboration network.
The resulting team includes ``Jure Leskovec," ``Jiawei Han," and ``Marko Grobelnik," as depicted in Figure \ref{fig:teamformation1}, with team members in blue and other nodes in gray.  Suppose we want to explain why ``Rayid Ghani'', shown with an orange circle, was excluded.  A skill-based counterfactual explanation is shown in Figure \ref{fig:teamformation2}, where adding the skills ``community'' and ``discovery'' puts Ghani on the team instead of Grobelnik.


\section{Experiments} \label{sec:evaluation}

We now present experimental results, starting with the setup (Section~\ref{subsec:setup}), and moving to the effectiveness-efficiency tradeoff of our pruning strategies for expert search (Section~\ref{sec:experiments}) and team formation (Section~\ref{subsec:team}), parameter sensitivity analysis (Section~\ref{sec:parametersensitivityanalysis}), and examples and case studies   (Section~\ref{sec:casestudies}). 
ExES consists of a web app frontend, developed with VueJS, and a backend REST API, developed with Flask and Python 3.10.12. We deployed the frontend and backend server and ran our experiments on a Ubuntu virtual machine, with an Intel Core i9-7920X CPU, 128 GB of RAM, and a GeForce RTX 4090 GPU.

\subsection{Setup}
\label{subsec:setup}


We evaluate ExES on two well-known datasets: \textit{DBLP} and \textit{GitHub}, described in Table \ref{tab:datasets}.  In DBLP \cite{dblp}, the collaboration network comprises researchers as nodes and paper co-authorship as edges.  We extracted experts' skills from their paper titles and abstracts by taking the top-scoring keywords according to TF-IDF, for an average of 15 skills per expert.  The GitHub dataset includes GitHub users as nodes and project collaborations as edges.  We extracted skills by applying a simlar TF-IDF methodology on the  descriptions and tags of user repositories. 
\begin{table}
\small
    \centering
       \caption{Dataset statistics} 
    \begin{tabular}{cccc}
    \hline
        \textbf{Dataset} & \# \textbf{Nodes} & \# \textbf{Edges} & \# \textbf{Skills}  \\
    \hline
        DBLP & 17630 & 128809 & 1829 \\
        GitHub & 3278 & 15502 & 863 \\
        \hline
    \end{tabular}
    \label{tab:datasets}
\end{table}

For each dataset, we generate 100 random queries, by sampling between 3 and 5 keywords uniformly from the universe of skills ($S$) of the corresponding dataset.  We then compare ExES's explanations against those generated by exhaustive search without any pruning, in terms of efficiency and effectiveness. We set a timeout of 1000 seconds to make sure the experiments finish in reasonable time. 

For skill addition counterfactuals (recall Section \ref{counterfactualskill}), exhaustive search probes the underlying expert search model by adding every skill to each node of the collaboration network. This makes it infeasible to run the exhaustive baseline.  Instead, we conducted two separate baseline tests. The first, denoted \mbox{\textit{Exhaustive neighborhood}} baseline, utilizes the entire network for potential node perturbations while using the pruned skill set from ExES as the candidate skills for addition. The metrics for this baseline are represented as \textit{N} in our tables.
The second baseline, denoted \mbox{\textit{Exhaustive skills}}, uses the full universe of skills ($S$) as potential additional skills but limits modifications to the neighborhood of the target expert. The results corresponding to this baseline are denoted as \textit{S} in our tables.

By default, we set beam size $b=30$, maximum explanation size $\gamma = 5$, number of required explanations $e=5$, and number of candidate features $t=10$. In addition, we set the neighborhood distance threshold $d$ to one for skill factuals, skill counterfactuals, and collaboration addition counterfactuals, ensuring an individual's neighborhood contains themselves and their immediate collaborators. For collaboration factuals and collaboration removal counterfactuals, we extended the neighborhood depth to $2$ to incorporate 2-hop collaborations.
We set $\tau$ in calculating collaboration SHAP values to $0.1$. In Section \ref{sec:parametersensitivityanalysis}, we will discuss how variations in these parameters affect the performance and outcomes of ExES.

We measure latency as the time required to calculate all explanations for a given individual.  This encompasses calculating feature attributions for every input feature in factual explanations (recall Section \ref{saliency}), and generating the top $e$ minimal explanations in counterfactual scenarios (recall Section \ref{counterfactual}).  In terms of the effectiveness of factual explanations, we check whether the important features found by ExES are similar to those found by exhaustive search.  To measure this, we utilize the \textbf{Precision@k} metric. Given a set of factual explanations generated by ExES, this measures the proportion of the top-$k$ important features (ranked by their SHAP scores) that also receive non-zero scores in the corresponding exhaustive search explanations.  For counterfactual explanations, we define
Precision as the fraction of explanations found by ExES that have the same minimal size as those found by exhaustive search.  Furthermore, we define Precision* as the fraction of explanations found by ExES that are either minimal or within one perturbation away from minimal. 

\subsection{Effect of Pruning - Expert Search} \label{sec:experiments}


For this set of experiments, we implemented an expert search model that uses Graph Convolutional Neural Networks and combines ideas from several state-of-the-art solutions \cite{hamidi2023variational,teamnegativesampling,hao2021ks}. 
We first ranked experts for each query using our pre-trained model, with $k=10$. Then, from the retrieved individuals, we sampled 100 experts within the top-$k$ and 100 non-experts ranked between $k+1$ and $2k$. We applied all factual and counterfactual explanation methods for each sampled individual to explain their ranking.

\begin{table}[t]
\small
\centering
    \caption{Factual explanation results: expert search}
\label{tab:timemetricsfactual}
\begin{tabular}{cccccc}
\hline
\multirow{2}{*}{\textbf{Features}} & \multirow{2}{*}{\textbf{Dataset}} & \multicolumn{2}{c}{\textbf{Latency (s)}} & \multicolumn{2}{c}{\textbf{Explanation Size}} \\ \cline{3-6} 
 &  & ExES & Baseline & ExES & Baseline \\ \hline
\multirow{2}{*}{Skills} & DBLP & 3.17 & 147.77 & 22.4 & 97.30 \\
 & GitHub & 1.11 & 147.22 & 21.6 & 43.82 \\ \hline
\multirow{2}{*}{Query terms} & DBLP & 0.13 & \multirow{2}{*}{---} & 2.76 & \multirow{2}{*}{---} \\
 & GitHub & 0.19 &  & 3.52 &  \\ \hline
\multirow{2}{*}{Collaborations} & DBLP & 11.29 & 98.02 & 13.17 & 174.96 \\
 & GitHub & 7.23 & 76.57 & 4.24 & 144.78
\end{tabular}
\end{table}
\begin{table}[t]
\small
\centering
    \caption{Counterfactual explanation resuls: expert search}
\label{tab:timemetricscounterfactual}
\begin{tabular}{cccccc}
\hline
\multirow{2}{*}{\textbf{Method}} & \multirow{2}{*}{\textbf{Dataset}} & \multicolumn{2}{c}{\textbf{Latency (s)}} & \multicolumn{2}{c}{\textbf{Explanation Size}} \\ \cline{3-6} 
 &  & ExES & Baseline & ExES & Baseline \\ \hline
\multirow{2}{*}{\begin{tabular}[c]{@{}c@{}}Skill Removal\\ (Experts)\end{tabular}} & DBLP & 57.53 & 917.06 & 2.23 & 1.53 \\
 & GitHub & 5.74 & 14.93 & 1.68 & 1.36 \\ \hline
\multirow{2}{*}{\begin{tabular}[c]{@{}c@{}}Query Augment.\\ (Experts)\end{tabular}} & DBLP & 0.36 & 0.21 & 1.17 & 1.00 \\
 & GitHub & 0.35 & 0.13 & 1.21 & 1.00 \\ \hline
\multirow{2}{*}{\begin{tabular}[c]{@{}c@{}}Link Removal\\ (Experts)\end{tabular}} & DBLP & 17.18 & 671.95 & 2.09 & 1.73 \\
 & GitHub & 12.25 & 154.25 & 2.31 & 2.23 \\ \hline
\multirow{4}{*}{\begin{tabular}[c]{@{}c@{}}Skill Addition\\ (Non-experts)\end{tabular}} & \multirow{2}{*}{DBLP} & \multirow{2}{*}{79.92} & N: 213.27 & \multirow{2}{*}{1.97} & N: 1.41 \\
 &  &  & S: 173.83 &  & S: 1.22 \\
 & \multirow{2}{*}{GitHub} & \multirow{2}{*}{5.16} & N: 15.24 & \multirow{2}{*}{2.21} & N: 1.63 \\
 &  &  & S: 13.19 &  & S: 1.18 \\ \hline
\multirow{2}{*}{\begin{tabular}[c]{@{}c@{}}Query Augment.\\ (Non-experts)\end{tabular}} & DBLP & 0.71 & 0.93 & 1.35 & 1.00 \\
 & GitHub & 0.51 & 0.64 & 1.06 & 1.00 \\ \hline
\multirow{2}{*}{\begin{tabular}[c]{@{}c@{}}Link Addition\\ (Non-experts)\end{tabular}} & DBLP & 6.17 & 159.79 & 1.33 & 1.12 \\
 & GitHub & 1.51 & 11.82 & 1.54 & 1.03 \\ \hline
\end{tabular}
\end{table}

\begin{table}[t]
\small
\centering
    \caption{Factual explanation precision: expert search}
\label{tab:saliencyprecision}
\begin{tabular}{cccc}
\hline
\textbf{Features}&\textbf{Dataset}& \textbf{Precision@1}&\textbf{Precision@5}\\ 
 \hline
  \multirow{2}{*}{Skills}&DBLP& 0.85&0.70\\
 & GitHub& 0.81&0.64\\
 \hline
 \multirow{2}{*}{Collaborations}&DBLP& 1.00&0.98\\
 & GitHub& 1.00&0.99\\
\hline
 
\end{tabular}
\end{table}
\begin{table}[t]
\small
\centering
    \caption{Counterfactual explanation precision: expert search}
\label{tab:accuracy}
\begin{tabular}{cccccc}
\hline
\multirow{2}{*}{\textbf{Explanation}} & \multirow{2}{*}{\textbf{Dataset}} & \multicolumn{2}{c}{\textbf{\# Explanations}} & \multirow{2}{*}{\textbf{Prec.}} & \multirow{2}{*}{\textbf{Prec.*}} \\ \cline{3-4}
 &  & ExES & Baseline &  &  \\ \hline
\multirow{2}{*}{\begin{tabular}[c]{@{}c@{}}Skill Removal\\ (Experts)\end{tabular}} & DBLP & 383 & 265 & 0.87 & 0.98 \\
 & GitHub & 465 & 385 & 0.98 & 0.99 \\ \hline
\multirow{2}{*}{\begin{tabular}[c]{@{}c@{}}Query Augment.\\ (Experts)\end{tabular}} & DBLP & 470 & 470 & 0.85 & 0.97 \\
 & GitHub & 435 & 475 & 0.81 & 0.96 \\ \hline
\multirow{2}{*}{\begin{tabular}[c]{@{}c@{}}Link Removal\\ (Experts)\end{tabular}} & DBLP & 301 & 230 & 0.93 & 1.00 \\
 & GitHub & 242 & 226 & 0.83 & 0.98 \\ \hline
\multirow{4}{*}{\begin{tabular}[c]{@{}c@{}}Skill Addition\\ (Non-experts)\end{tabular}} & \multirow{2}{*}{DBLP} & \multirow{2}{*}{387} & N: 270 & N: 0.87 & N: 0.97 \\
 &  &  & S: 420 & S: 0.43 & S: 0.85 \\
 & \multirow{2}{*}{GitHub} & \multirow{2}{*}{375} & N: 295 & N: 0.91 & N: 0.96 \\
 &  &  & S: 440 & S: 0.32 & S: 0.73 \\ \hline
\multirow{2}{*}{\begin{tabular}[c]{@{}c@{}}Query Augment.\\ (Non-experts)\end{tabular}} & DBLP & 456 & 460 & 0.59 & 0.91 \\
 & GitHub & 440 & 475 & 0.64 & 0.94 \\ \hline
\multirow{2}{*}{\begin{tabular}[c]{@{}c@{}}Link Addition\\ (Non-experts)\end{tabular}} & DBLP & 441 & 460 & 0.73 & 0.95 \\
 & GitHub & 400 & 470 & 0.59 & 0.89 \\ \hline
\end{tabular}
\end{table}

Tables \ref{tab:timemetricsfactual} and \ref{tab:timemetricscounterfactual} show the average latency and explanation sizes (i.e., the number of features with non-zero SHAP values) for ExES and the exhaustive baseline for DBLP and GitHub. Since the exhaustive baseline takes every feature into account, factual exhaustive explanations can be larger than ours. 

Furthermore, we see that ExES is often more than an order of magnitude faster, except in counterfactual query explanations for turning experts into non-experts.  This is likely because turning experts into non-experts can be done simply by adding random keywords to the query.  That said, the value of using keyword embeddings in ExES is to ensure that the counterfactual queries are close to the original queries.   

Table \ref{tab:saliencyprecision} shows Precision@1 and Precision@5 for skill and collaboration factual explanations.  We see that ExES often finds the same explanations as exhaustive search, but often an order of magnitude faster (recall Table \ref{tab:timemetricsfactual}).  
Next, Table \ref{tab:accuracy} shows the number of counterfactual explanations generated by ExES and by exhaustive search within the time limit, along with the Precision and Precision* of the explanations found by ExES.
Given a total of 100 target experts and $e = 5$, the ideal output would involve 500 explanations for each method. However, ExES may miss some explanations due to pruning of the search space, while exhaustive search might not complete within the time limit.
ExES has an average precision of $74\%$ over all counterfactual explanation types. In addition, ExES has an average precision* of $94\%$, i.e., it  generates \textit{nearly-optimal} explanations in $94\%$ of the cases.  In Section~\ref{sec:casestudies}, we show examples where the explanations generated by ExES may not be as concise as those found by exhaustive search but are still meaningful and actionable.

For skill addition counterfactuals, the precision of ExES drops below $0.5$. Thus, selecting appropriate candidate keywords is crucial.  One way to increase precision is to increase $t$, at the expense of latency.  A larger $t$ value leads to ExES behaving more closely to the \textit{Exhaustive skillset} baseline, but at the cost of increased computation time. Still, the precision results indicate that $70\%$ ExES's skill addition explanations are nearly optimal.

\subsection{Effect of Pruning - Team Formation}
\label{subsec:team}

\begin{table}[t]
\small
\centering
    \caption{Factual explanation results: team formation}
\label{tab:teamtimemetricsfactual}
\begin{tabular}{cccccc}
\hline
\multirow{2}{*}{\textbf{Features}} & \multirow{2}{*}{\textbf{Dataset}} & \multicolumn{2}{c}{\textbf{Latency (s)}} & \multicolumn{2}{c}{\textbf{Explanation Size}} \\ \cline{3-6} 
 &  & ExES & Baseline & ExES & Baseline \\ \hline
\multirow{2}{*}{Skills} & DBLP & 8.63 & 236.53 & 30.65 & 93.40 \\
 & GitHub & 2.88 & 229.80 & 23.98 & 40.69 \\ \hline
\multirow{2}{*}{Query terms} & DBLP & 0.24 & \multirow{2}{*}{---} & 2.65 & \multirow{2}{*}{---} \\
 & GitHub & 0.31 &  & 3.52 &  \\ \hline
\multirow{2}{*}{Collaborations} & DBLP & 62.41 & 645.62 & 18.38 & 209.38 \\
 & GitHub & 41.02 & 79.04 & 8.62 & 168.73
\end{tabular}
\end{table}
\begin{table}[t]
\small
\centering
    \caption{Counterfactual explanation results: team formation}
\label{tab:teamtimemetricscounterfactual}
\begin{tabular}{cccccc}
\hline
\multirow{2}{*}{\textbf{Explanation}} & \multirow{2}{*}{\textbf{Dataset}} & \multicolumn{2}{c}{\textbf{Latency (s)}} & \multicolumn{2}{c}{\textbf{Explanation Size}} \\ \cline{3-6} 
 &  & ExES & Baseline & ExES & Baseline \\ \hline
\multirow{2}{*}{\begin{tabular}[c]{@{}c@{}}Skill Removal\\ (Members)\end{tabular}} & DBLP & 132.18 & 958.05 & 1.77 & 1.31 \\
 & GitHub & 10.12 & 31.82 & 1.54 & 1.26 \\ \hline
\multirow{2}{*}{\begin{tabular}[c]{@{}c@{}}Query Augment.\\ (Members)\end{tabular}} & DBLP & 0.82 & 0.46 & 1.20 & 1.05 \\
 & GitHub & 0.80 & 0.35 & 1.30 & 1.00 \\ \hline
\multirow{2}{*}{\begin{tabular}[c]{@{}c@{}}Link Removal\\ (Members)\end{tabular}} & DBLP & 35.91 & 932.63 & 1.95 & 1.60 \\
 & GitHub & 22.74 & 328.47 & 2.08 & 1.64 \\ \hline
\multirow{4}{*}{\begin{tabular}[c]{@{}c@{}}Skill Addition\\ (Non-members)\end{tabular}} & \multirow{2}{*}{DBLP} & \multirow{2}{*}{181.46} & N: 450.35 & \multirow{2}{*}{2.19} & N: 1.59 \\
 &  &  & S: 371.57 &  & S: 1.31 \\
 & \multirow{2}{*}{GitHub} & \multirow{2}{*}{13.78} & N: 28.40 & \multirow{2}{*}{2.88} & N: 1.95 \\
 &  &  & S: 24.72 &  & S: 1.58 \\ \hline
\multirow{2}{*}{\begin{tabular}[c]{@{}c@{}}Query Augment.\\ (Non-members)\end{tabular}} & DBLP & 0.84 & 1.28 & 2.14 & 1.83 \\
 & GitHub & 0.59 & 0.75 & 2.03 & 1.67 \\ \hline
\multirow{2}{*}{\begin{tabular}[c]{@{}c@{}}Link Addition\\ (Non-members)\end{tabular}} & DBLP & 13.32 & 366.51 & 1.51 & 1.26 \\
 & GitHub & 4.26 & 25.33 & 1.79 & 1.15 \\ \hline
\end{tabular}
\end{table}
\begin{table}[t]
\small
\centering
    \caption{Factual explanation precision: team formation}
\label{tab:teamsaliencyprecision}
\begin{tabular}{cccc}
\hline
\textbf{Features}&\textbf{Dataset}& \textbf{Precision@1}&\textbf{Precision@5}\\ 
 \hline
 \multirow{2}{*}{Skills}&DBLP& 0.77&0.57\\
 & GitHub& 0.58&0.51\\
 \midrule
 \multirow{2}{*}{Collaborations}&DBLP& 1.00&0.89\\
 & GitHub& 1.00&0.91\\
\hline
 
\end{tabular}
\end{table}
\begin{table}[t]
\small
\centering
    \caption{Counterfact. explanation precision: team formation}
\label{tab:teamaccuracy}
\begin{tabular}{cccccc}
\hline
\multirow{2}{*}{\textbf{Explanation}} & \multirow{2}{*}{\textbf{Dataset}} & \multicolumn{2}{c}{\textbf{\# Explanations}} & \multirow{2}{*}{\textbf{Prec.}} & \multirow{2}{*}{\textbf{Prec.*}} \\ \cline{3-4}
 &  & ExES & Exhaustive &  &  \\ \hline
\multirow{2}{*}{\begin{tabular}[c]{@{}c@{}}Skill Removal\\ (Members)\end{tabular}} & DBLP & 306 & 210 & 0.82 & 0.92 \\
 & GitHub & 405 & 365 & 0.92 & 0.93 \\ \hline
\multirow{2}{*}{\begin{tabular}[c]{@{}c@{}}Query Augment.\\ (Members)\end{tabular}} & DBLP & 260 & 260 & 0.86 & 0.98 \\
 & GitHub & 199 & 250 & 0.82 & 0.97 \\ \hline
\multirow{2}{*}{\begin{tabular}[c]{@{}c@{}}Link Removal\\ (Members)\end{tabular}} & DBLP & 296 & 225 & 0.92 & 1.00 \\
 & GitHub & 223 & 215 & 0.82 & 0.97 \\ \hline
\multirow{4}{*}{\begin{tabular}[c]{@{}c@{}}Skill Addition\\ (Non-members)\end{tabular}} & \multirow{2}{*}{DBLP} & \multirow{2}{*}{350} & N: 250 & N: 0.81 & N: 0.96 \\
 &  &  & S: 405 & S: 0.41 & S: 0.82 \\
 & \multirow{2}{*}{GitHub} & \multirow{2}{*}{348} & N: 265 & N: 0.90 & N: 0.97 \\
 &  &  & S: 410 & S: 0.30 & S: 0.70 \\ \hline
\multirow{2}{*}{\begin{tabular}[c]{@{}c@{}}Query Augment.\\ (Non-members)\end{tabular}} & DBLP & 425 & 445 & 0.50 & 0.90 \\
 & GitHub & 410 & 455 & 0.60 & 1.00 \\ \hline
\multirow{2}{*}{\begin{tabular}[c]{@{}c@{}}Link Addition\\ (Non-members)\end{tabular}} & DBLP & 410 & 445 & 0.69 & 0.91 \\
 & GitHub & 360 & 450 & 0.57 & 0.83 \\ \hline
\end{tabular}
\end{table}

For this set of experiments, we use the recently proposed team formation method from \cite{hao2021ks}. This method requires the user to input an expert as the main team member, and constructs a team around the main member until all the query terms are covered.  
We used the same queries and parameter values as in the previous set of experiments for expert search.  For each query, we randomly selected an expert from the top-$k$ list and then used the above team formation method to build a team around that expert. Within each formed team, we randomly sampled one team member to explain their inclusion and one non-member from the main member's neighborhood to explain their exclusion.

Tables \ref{tab:teamtimemetricsfactual} and \ref{tab:teamtimemetricscounterfactual} compare the average latency and explanation sizes using ExES and the exhaustive baseline.  These results are similar to those for expert search: in many cases, ExES is an order of magnitude or more faster.  
Next, Table \ref{tab:teamsaliencyprecision} shows the Precision@1 and Precision@5 scores for skill and collaboration factual explanations generated by ExES.  We see that ExES achieves a Precision@1 of 0.83 and a Precision@5 of 0.72 on average.
Furthermore, Table \ref{tab:teamaccuracy} displays the number of explanations generated by ExES and by the baseline within the time limit, along with the Precision and Precision* of the explanations created by ExES. Similar to our approach for evaluating expert search explanations, \textit{Exhaustive neighborhood} and \textit{Exhaustive skillset} baselines are employed for evaluating skill addition counterfactuals. The results indicate that ExES has an average Precision of $71\%$ and an average Precision* of $91\%$ over all the counterfactual explanation types.

\subsection{Parameter Sensitivity} \label{sec:parametersensitivityanalysis}

We now explore the impact of the beam size $b$, the number of candidate features selected in beam search $t$, the neighborhood radius $d$, and the threshold for calculating collaboration SHAP values $\tau$. We evaluate their impacts on latency, precision, the number of explanations found, and explanation size. To analyze each parameter, we set the other parameters to the default values mentioned earlier.  We show several representative plots below, with other combinations of parameters and explanation types showing similar trends.

\begin{figure*}
    \centering
    \begin{subfigure}[b]{0.24\textwidth}
        \centering
        \includegraphics[width=\textwidth]{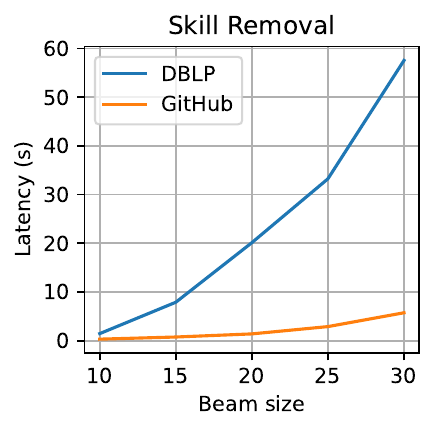}
        \caption{Effect of $b$ on Latency}
        \label{fig:sub1lat}
    \end{subfigure}
    \begin{subfigure}[b]{0.24\textwidth}
        \centering
        \includegraphics[width=\textwidth]{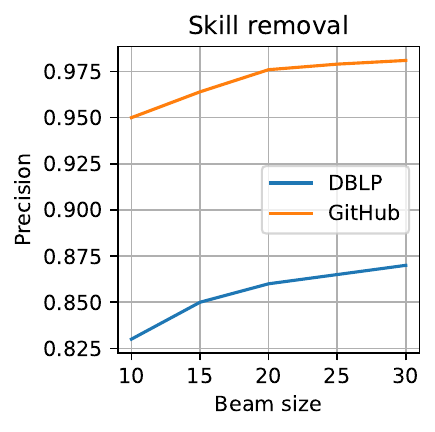}
        \caption{Effect of $b$ on Precision}
        \label{fig:sub1pre}
    \end{subfigure}
    \hfill 
    \begin{subfigure}[b]{0.24\textwidth}
        \centering
        \includegraphics[width=\textwidth]{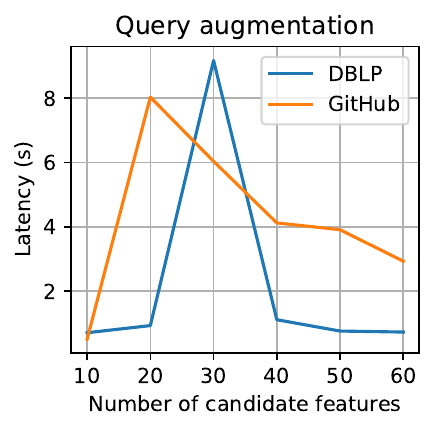}
        \caption{Effect of $t$ on Latency}
        \label{fig:sub2lat}
    \end{subfigure}
    \begin{subfigure}[b]{0.24\textwidth}
        \centering
        \includegraphics[width=\textwidth]{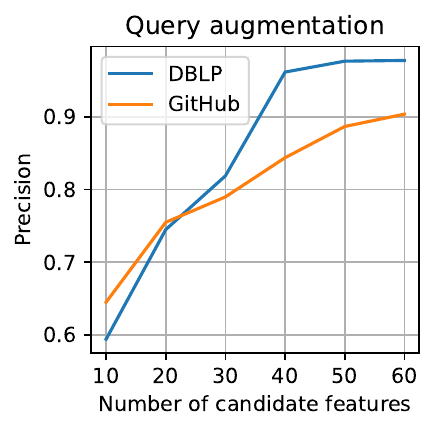}
        \caption{Effect of $t$ on Precision}
        \label{fig:sub2pre}
    \end{subfigure}
    \hfill 
    \begin{subfigure}[b]{0.24\textwidth}
        \centering
        \includegraphics[width=\textwidth]{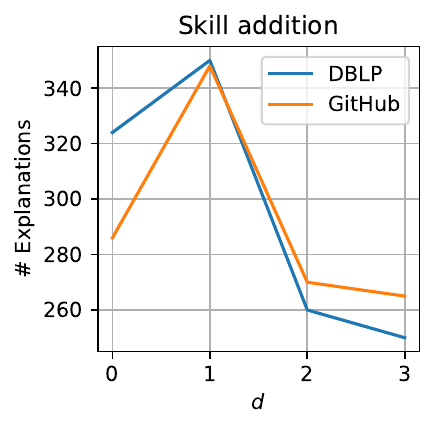}
        \caption{Effect of $d$ on \# Expl.}
        \label{fig:sub4}
    \end{subfigure}
    \begin{subfigure}[b]{0.24\textwidth}
        \centering
        \includegraphics[width=\textwidth]{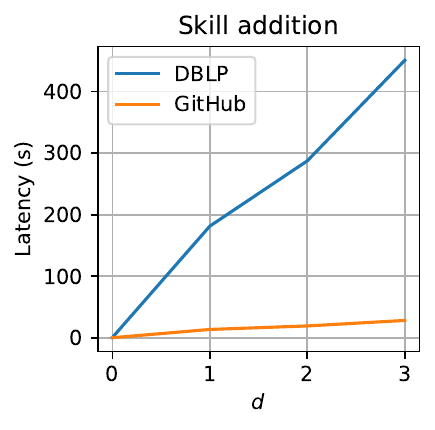}
        \caption{Effect of $d$ on Latency}
        \label{fig:sub4lat}
    \end{subfigure}
    \begin{subfigure}[b]{0.24\textwidth}
        \centering
        \includegraphics[width=\textwidth]{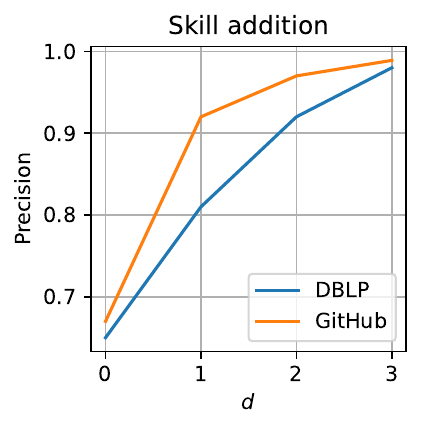}
        \caption{Effect of $d$ on Precision}
        \label{fig:sub4pre}
    \end{subfigure}
    \begin{subfigure}[b]{0.24\textwidth}
        \centering
        \includegraphics[width=\textwidth]{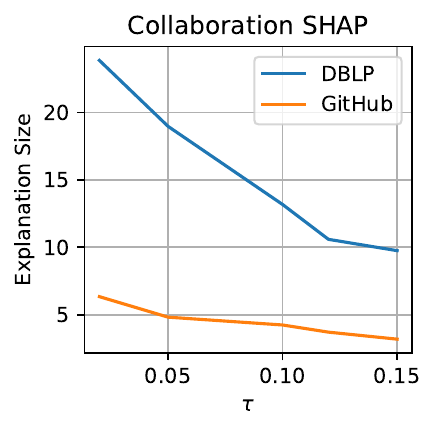}
        \caption{Effect of $\tau$ on Expl. Size}
        \label{fig:sub3}
    \end{subfigure}
    
    \caption{Parameter sensitivity analysis}
    \label{fig:parameterplot}
\end{figure*}

Figures \ref{fig:sub1lat} and \ref{fig:sub1pre} show the effect of beam size $b$ on the latency and precision of skill removal explanations in expert search: as $b$ increases, so does the running time and precision as a result of widening the search space. 
Figures \ref{fig:sub2lat} and \ref{fig:sub2pre} measure the effect of the number of candidate tokens $t$ on the latency and precision of query augmentation explanations for non-experts in expert search. Figure \ref{fig:sub2lat} shows an increasing trend in latency for lower values of $t$. However, for larger values of $t$, the set of candidate additional keywords grows, increasing the likelihood of finding effective keywords for counterfactuals. This allows the search to terminate earlier.
Figure \ref{fig:sub2pre} verifies that increasing $t$ increases precision, but the gains decrease for large values of $t$.  

Next, Figure \ref{fig:sub4} demonstrates the impact of neighborhood size $d$ on the number of skill addition explanations in expert search. In this figure, smaller values of $d$ result in ExES failing to find valid explanations due to the restricted search space, while larger values of $d$ cause ExES to exceed the time limit. Moreover, Figures \ref{fig:sub4lat} and \ref{fig:sub4pre} show the tradeoff between latency and precision (compared with the \textit{Exhaustive neighborhood} baseline) for different values of $d$.
Finally, Figure \ref{fig:sub3} shows the effect of the threshold $\tau$ on explanation size in collaboration SHAP values in expert search. Increasing $\tau$ leads to smaller explanation since fewer experts are considered ``impactful" (recall Pruning Strategy 2 from Section~\ref{saliency}).

\subsection{Case Studies}
\label{sec:casestudies}

In this section, we explore the functionality of ExES in explaining expert search and team formation outcomes.  We also show examples where ExES finds less concise explanations than exhaustive search, but those explanations are still meaningful.  

First, we run the query ``deep neural training" on the DBLP collaboration network with $k=10$, to factually explain ``Yann LeCun". 
\begin{figure}
    \centering
    \begin{subfigure}[b]{0.9\linewidth}
        \centering
        \includegraphics[width=\linewidth]{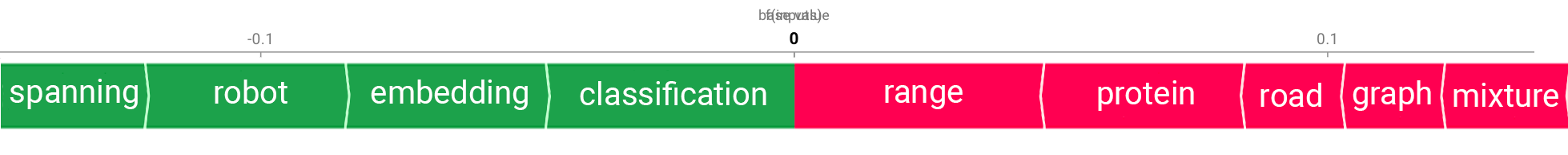}
        \caption{With pruning}
        \label{fig:shaplecun1}
    \end{subfigure}
    \hfill 
    \begin{subfigure}[b]{0.9\linewidth}
        \centering
        \includegraphics[width=\linewidth]{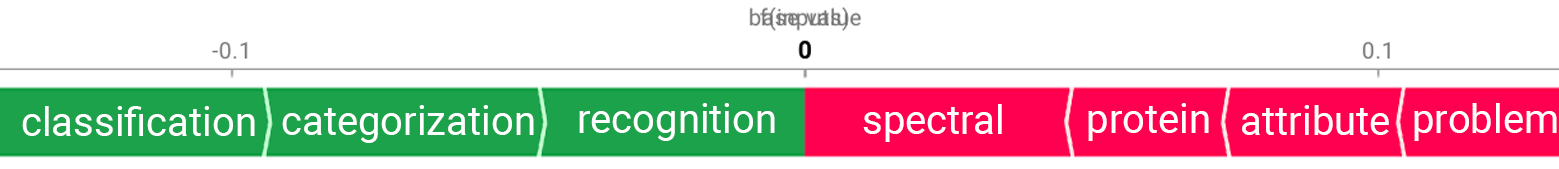}
        \caption{Without pruning}
        \label{fig:shaplecun_baseline}
    \end{subfigure}
    \caption{Skill SHAP values for LeCun}
    \label{fig:shaplecun}
\end{figure}
Figure \ref{fig:shaplecun} shows the skill SHAP values for ExES (top) and exhaustive search (bottom). The influential skills found be ExES are related to machine learning, such as ``classification" and ``embedding". On the other hand, application-related skills such as ``protein'' are not relevant.  Exhaustive search finds similar patterns even though some of the specific skills it finds relevant and irrelevant are different.  
\begin{figure}
    \centering
    \includegraphics[width=0.99\linewidth]{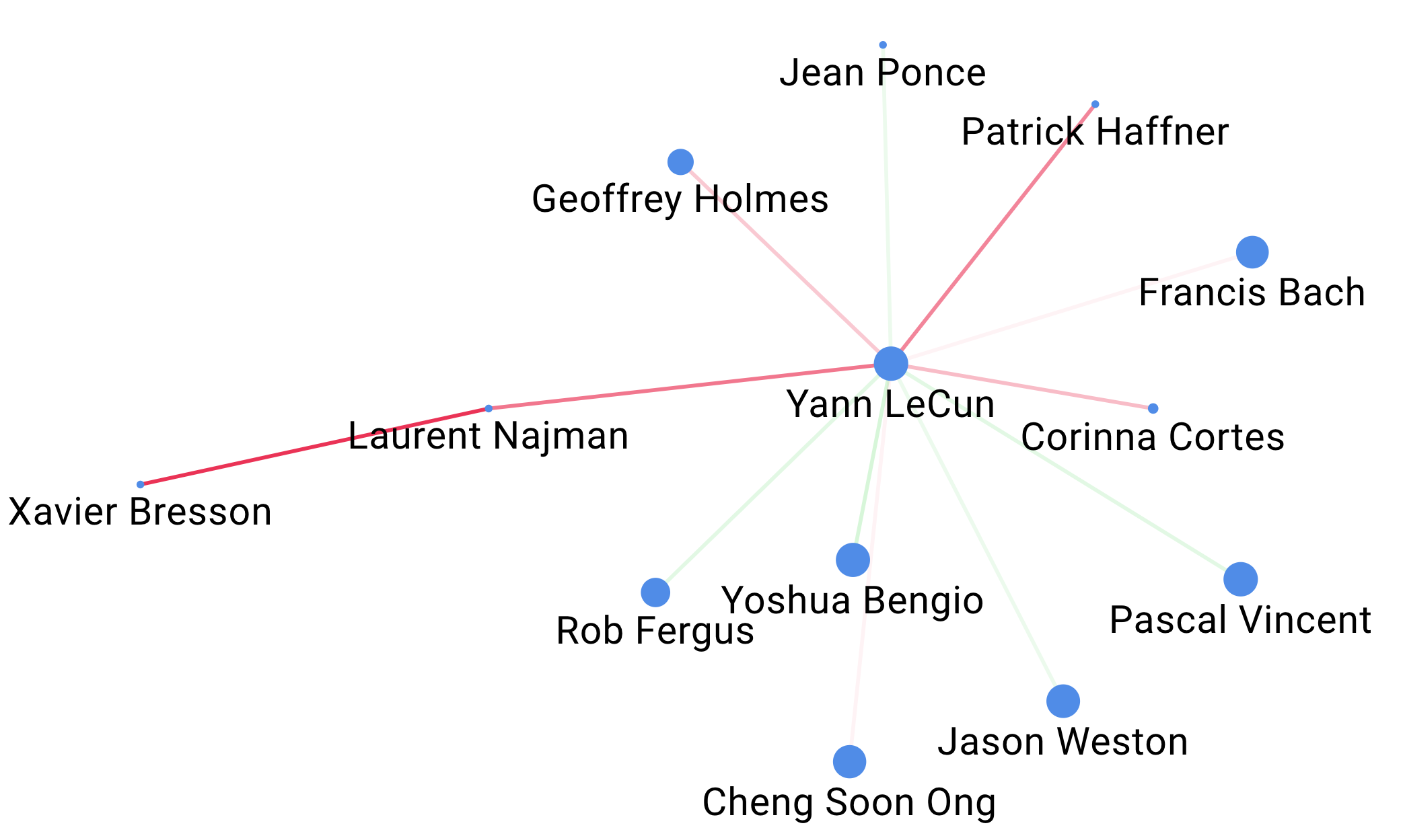}
    \caption{Collaboration SHAP values for LeCun}
    \label{fig:shapedgelecun}
\end{figure}
Next, we analyze the SHAP values of collaborations in LeCun's neighborhood, shown in Figure \ref{fig:shapedgelecun}
Collaborations with ``Yoshua Bengio", ``Pascal Vincent", and ``Rob Fergus", all established deep learning researchers, appear to have a positive impact. Other collaborators, such as ``Laurent Najman", ``Xavier Bresson", and ``Patrick Haffner", have a negative impact with respect to this query, likely because they are experts in other machine learning subfields. 

Next, we take the same query and $k$, but counterfactually explain why Bengio placed outside the top-10.
\begin{figure}
    \centering
    \includegraphics[width=\linewidth]{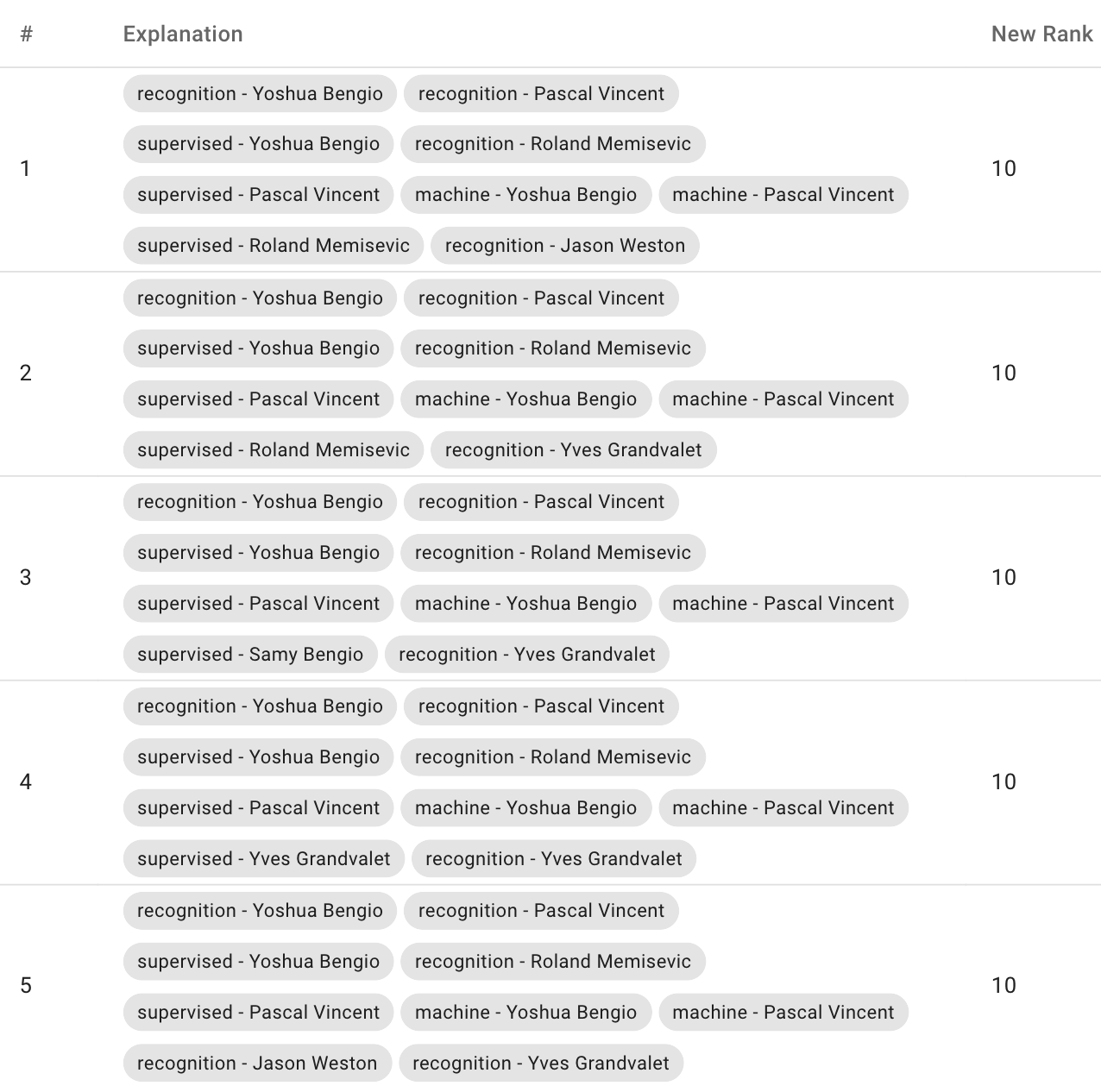}
    \caption{Counterfactual skill explanations for Bengio}
    \label{fig:bengiocf1}
\end{figure}
Figure \ref{fig:bengiocf1} shows a list of minimal counterfactual skill explanations of size nine, i.e., nine skills would have to be added to place Bengio in the top-10.  ExES suggests new skills for both Bengio and his network, which demonstrates the importance of neighborhood structure in expert search. 
\begin{figure}
    \centering
    \begin{subfigure}[b]{0.3\textwidth}
        \centering
        \includegraphics[width=\textwidth]{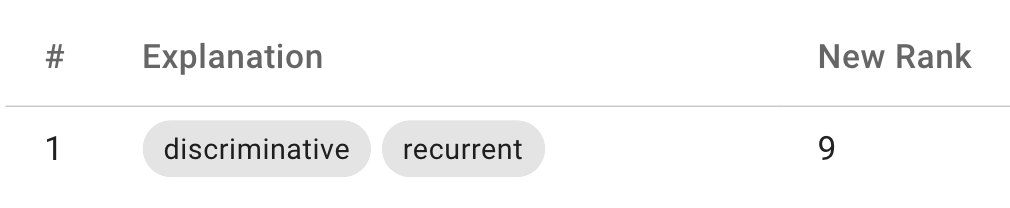}
        \caption{With pruning}
        \label{fig:bengiocf2p}
    \end{subfigure}
    \hfill 
    \begin{subfigure}[b]{0.3\textwidth}
        \centering        \includegraphics[width=\textwidth]{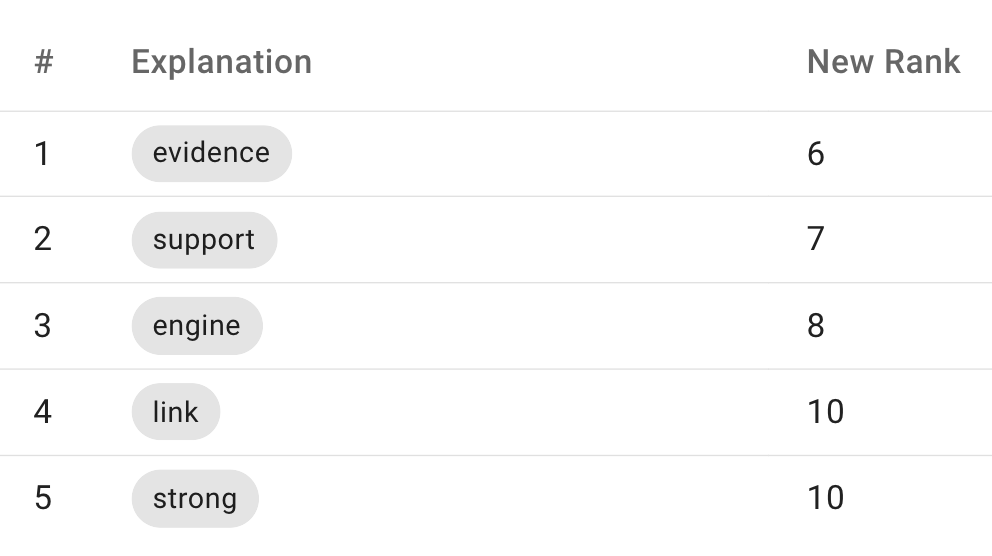}
        \caption{Without pruning}
        \label{fig:bengiocf2np}
    \end{subfigure}
    \caption{Counterfactual query explanations for Bengio}
    \label{fig:bengiocf2}
\end{figure}

\begin{figure}
    \centering
    \begin{subfigure}[b]{0.71\linewidth}
        \centering
        \includegraphics[width=\linewidth]{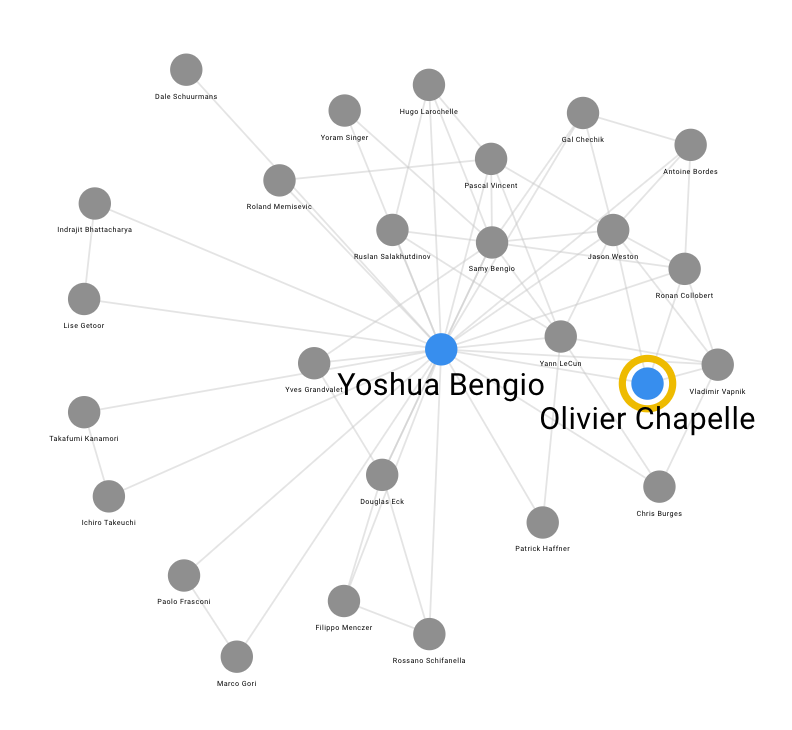}
        \caption{Original team}
        \label{fig:bengioteam1}
    \end{subfigure}
    \hfill 
    \begin{subfigure}[b]{\linewidth}
        \centering
        \includegraphics[width=\linewidth]{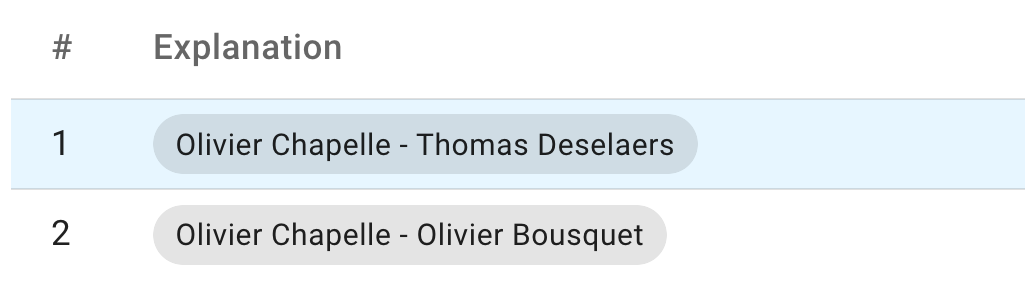}
        \caption{Counterfactual explanations}
        \label{fig:bengioteam2}
    \end{subfigure}
    \begin{subfigure}[b]{0.71\linewidth}
        \centering
        \includegraphics[width=\linewidth]{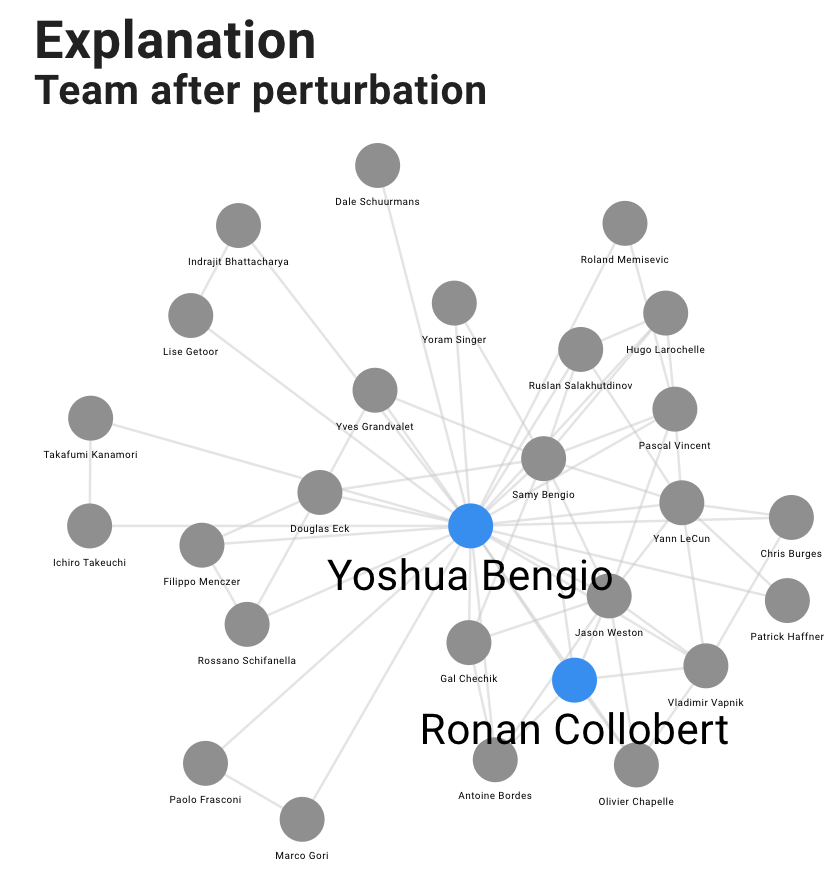}
        \caption{Modified team}
        \label{fig:bengioteam3}
    \end{subfigure}
    \caption{Team formation explanations}
    \label{fig:bengioteam}
\end{figure}

We now analyze counterfactual query explanations that put Bengio in the top-10. Figure \ref{fig:bengiocf2} compares the explanations generated by ExES (Figure \ref{fig:bengiocf2p}) with those by exhaustive search (Figure \ref{fig:bengiocf2np}).
ExES generates an explanation of size two (add ``discriminative" and ``recurrent" to the query), while the minimal counterfactuals identified by the exhaustive baseline have size one. The counterfactual query generated by ExES is relevant to the initial query (``discriminative" and ``recurrent" are terms associated with deep learning). In contrast, appending keywords such ``evidence" and ``support", as suggested by exhaustive search, does not seem meaningful.

Finally, we investigate team formation with the query ``deep neural training supervised", where our team formation model builds a team around Yoshua Bengio.  Olivier Chapelle is another team member, shown in Figure \ref{fig:bengioteam1}.  Figure \ref{fig:bengioteam2} shows a counterfactual collaboration explanation for Chapelle's inclusion in the team, i.e., two edge removals that would replace Chapelle with Ronan Collobert.  Focusing on the first edge removal, supervised learning is a skill associated with Deselaers, meaning that the loss of this collaborator would also affect Chapelle's ranking with respect to the given query, to the point that Chapelle would be replaced with Collobert, another expert in supervised learning.

\section{Conclusions} \label{sec:conclusion}

We introduced ExES, the first tool specifically designed to generate explanations for expert search and team formation systems. By framing these tasks as binary classification problems, we were able to apply existing factual and counterfactual explanation methods to this novel problem.  We also proposed pruning rules to speed up the explanation search.  One direction for future work is to extend ExES to other graph search domains such as keyword search in relational databases or protein interaction networks. Another extension could be to investigate explanation robustness: are similar individuals explained similarly in terms of their inclusion or exclusion in the list of top experts?  Furthermore, a user study could help identify practical applications of explainable expert search such as career advancement advice, as mentioned earlier.  Finally, an interesting area for future work is to explore the interplay between data quality and explanations \cite{hao2021ks}: 
the skills or connections suggested by ExES might correspond to missing elements in the underlying graph.     

\bibliographystyle{ACM-Reference-Format}
\bibliography{sample}

\end{document}